\documentclass[%
reprint, 
amsmath,amssymb,
aps,
pra,
 floatfix,
]{revtex4-1}

\usepackage{graphicx}
\usepackage{dcolumn}
\usepackage{bm}
\usepackage{physics}

\usepackage[usenames, dvipsnames]{color}
\usepackage{sidecap}
\usepackage{textcomp}
\usepackage{textgreek}
\usepackage{amsmath}

\begin{document}

\title{Coherence and entanglement of inherently long-lived spin pairs in diamond}

\author{H. P. Bartling$^{1,2}$}
\author{M. H. Abobeih$^{1,2}$}
\author{B. Pingault$^{1,2,3}$}
\author{M. J. Degen$^{1,2}$}
\author{S. J. H. Loenen$^{1,2}$}
\author{C. E. Bradley$^{1,2}$}
\author{J. Randall$^{1,2}$}
\author{M. Markham$^4$} 
\author{D. J. Twitchen$^4$}
\author{T. H. Taminiau$^{1,2}$}
\email{T.H.Taminiau@TUDelft.nl}

\affiliation{$^{1}$QuTech, Delft University of Technology, PO Box 5046, 2600 GA Delft, The Netherlands}
\affiliation{$^{2}$Kavli Institute of Nanoscience Delft, Delft University of Technology, PO Box 5046, 2600 GA Delft, The Netherlands}
\affiliation{$^{3}$John A. Paulson School of Engineering and Applied Sciences, Harvard University, 29 Oxford Street, Cambridge MA 02138, USA}
\affiliation{$^{4}$Element Six, Fermi Avenue, Harwell Oxford, Didcot, Oxfordshire, OX11 0QR, United Kingdom}

\date{\today}

\maketitle

\newpage

\newpage

\textbf{Understanding and protecting the coherence of individual quantum systems is a central challenge in quantum science and technology. Over the last decades, a rich variety of methods to extend coherence have been developed \cite{Miao2020, Bar-Gill2013, Muhonen2014, Bradley2019, Shulman2012, Shulman2014, Langer2005,Harty2014,Wang2017, Wolfowicz2013, Saeedi2013, Zhong2015, Bae2018, Abobeih2018}. A complementary approach is to look for naturally occurring systems that are inherently protected against decoherence. Here, we show that pairs of identical nuclear spins in solids form intrinsically long-lived quantum systems. We study three carbon-13 pairs in diamond \cite{Zhao2011, Shi2013, Abobeih2018} and realize high-fidelity measurements of their quantum states using a single NV center in their vicinity. We then reveal that the spin pairs are robust to external perturbations due to a unique combination of three phenomena: a clock transition, a decoherence-free subspace, and a variant on motional narrowing. The resulting inhomogeneous dephasing time is $\boldsymbol{T_2^* = 1.9(3)}$ minutes, the longest reported for individually controlled qubits \cite{Harty2014}. Finally, we develop complete control and realize an entangled state between two spin-pair qubits through projective parity measurements. These long-lived qubits are abundantly present in diamond and other solids, and provide new opportunities for quantum sensing \cite{Degen2017}, quantum information processing \cite{Bradley2019, Cramer2016,Waldherr2014}, and quantum networks \cite{Hensen2015}.}

Solid-state spins provide a versatile platform for investigating quantum physics and realizing quantum technologies \cite{Miao2020,Bar-Gill2013,Muhonen2014,Bradley2019,Shulman2012,Shulman2014,Wolfowicz2013,Saeedi2013,Zhong2015,Bae2018,Abobeih2018,Zhao2011,Shi2013,Degen2017,Cramer2016,Waldherr2014,Hensen2015,Reiserer2016,Seo2016,Ye2019,Liu2017,Kalb2016,Cujia2019,Pfender2019}. Various methods to extend coherence times have been developed for spin ensembles \cite{Bar-Gill2013,Zhong2015,Saeedi2013,Wolfowicz2013}, as well as individually controlled spin qubits \cite{Muhonen2014,Miao2020,Bradley2019,Abobeih2018, Shulman2012,Shulman2014, Waldherr2014,Cramer2016, Bae2018}. These methods include the precise tuning of magnetic fields to create magnetic-field insensitive clock transitions \cite{Wolfowicz2013,Harty2014,Bae2018,Langer2005, Zhong2015}, decoherence-free subspaces to protect against correlated noise \cite{Bae2018, Reiserer2016,Shulman2012,Langer2005}, dynamical decoupling to mitigate slowly varying noise \cite{Bar-Gill2013, Muhonen2014, Bradley2019, Zhong2015,Abobeih2018,Saeedi2013,Wang2017}, real-time Hamiltonian estimation \cite{Shulman2014}, quantum error correction \cite{Waldherr2014, Cramer2016}, and isotopic purification to remove the spin background \cite{Bar-Gill2013,Muhonen2014, Saeedi2013}.

To realize solid-state quantum systems that are inherently long-lived, we investigate pairs of identical interacting nuclear spins. Such spin pairs are naturally and abundantly present in solids like diamond, silicon, silicon-carbide, germanium, graphene and MoS$_2$ \cite{Zhao2011,Shi2013, Abobeih2018,Seo2016, Ye2019}. Traditionally, the dynamics of such spin pairs have been regarded as a primary noise source for solid-state spin qubits \cite{Zhao2012,Seo2016, Abobeih2018,Wolfowicz2013, Ye2019}. In contrast, we show that spin pairs themselves provide individually controllable and decoherence-protected quantum systems.

\begin{figure}
\includegraphics[width = \columnwidth]{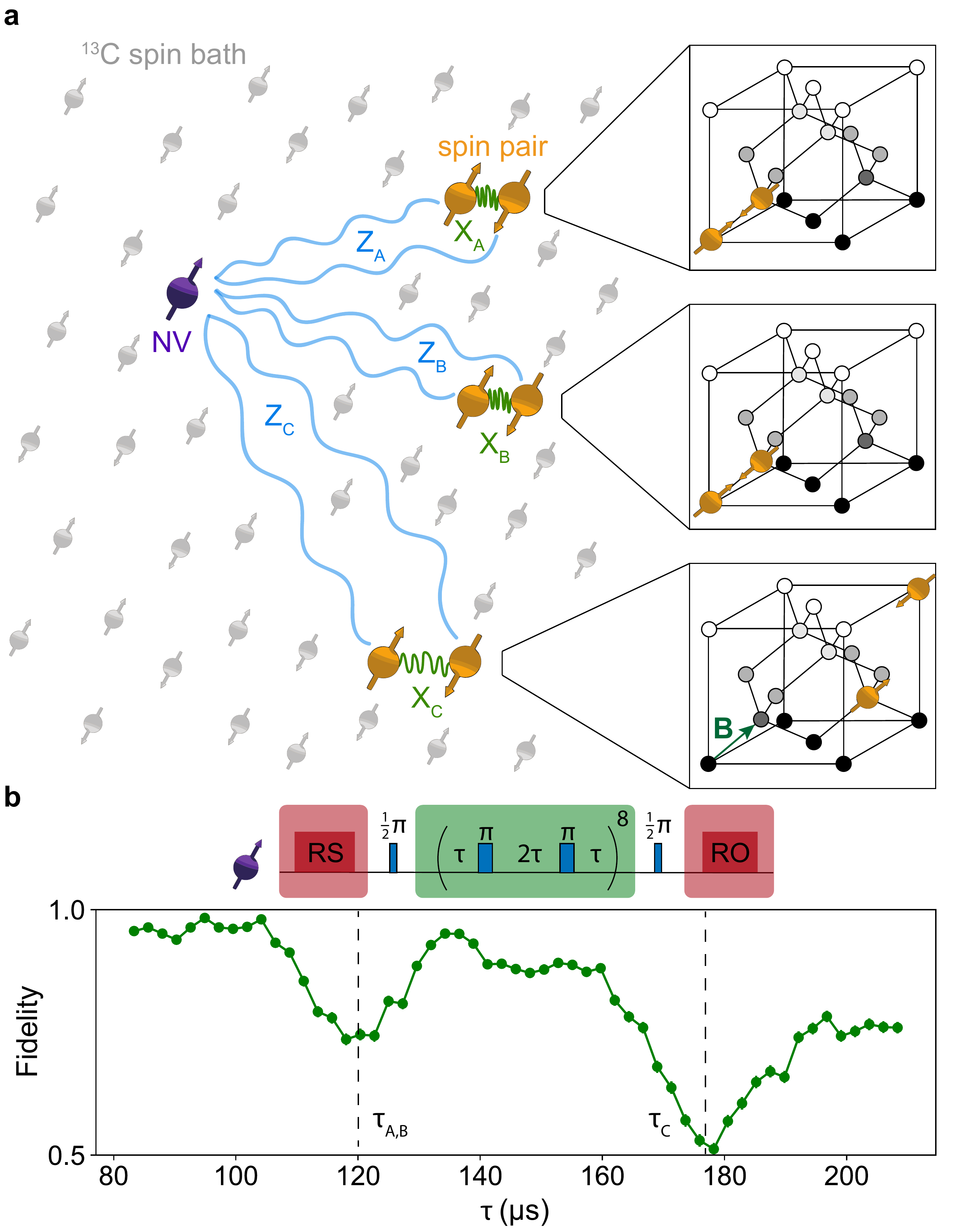}
\caption{ \textbf{System and basic spectroscopy.} \textbf{a.} We study three $^{13}$C spin pairs (A, B and C) in a diamond. The pairs are detected and controlled using a nearby NV center. The insets show the spatial configuration of the pairs. Pair A and B are nearest-neighbour pairs oriented along the external magnetic field $B_z$. For pair C we show one of the three possible orientations (Supplementary Section IV). The main source of decoherence is the surrounding bath of $^{13}$C spins ($1.1\%$ abundance). \textbf{b.} Sensing the pair pseudo-spins \cite{Zhao2011,Shi2013, Abobeih2018}. The NV electron spin is prepared in a superposition and a periodic sequence of $\pi$ pulses is applied. If the interpulse delay is resonant with the dynamics of a pair, a loss of electron coherence is observed. We set $\tau = m 2 \pi /\omega_L$ with $m$ an integer and $\omega_L$ the $^{13}$C Larmor frequency to avoid interactions with individual $^{13}$C spins \cite{Shi2013,Abobeih2018}. The vertical lines mark the values for $\tau$ used in this work for the three pairs ($\tau_A = \tau_B = 120.330$ \textmugreek s and $\tau_C = 177.026$ \textmugreek s). The NV spin is prepared (RS) and read out (RO) optically (see Methods).}
\label{Figure1}
\end{figure}

The system that we investigate is illustrated in Fig. 1a. We consider three pairs of coupled $^{13}$C nuclear spins in the vicinity of an NV center in a diamond at 3.7 K. The NV center provides a controllable electron spin with long coherence times that can be initialised and measured optically (Methods) \cite{Bar-Gill2013,Bradley2019,Waldherr2014,Cramer2016,Abobeih2018,Shi2013,Hensen2015}. Because the NV spin creates a switchable local magnetic-field gradient over each pair, it can be used to sense and manipulate the spin pairs \cite{Zhao2011,Shi2013,Abobeih2018}, despite their excellent protection from external influences.

A spin-1/2 pair is described by four states: $\ket{\uparrow\uparrow}$, $\ket{\uparrow\downarrow}$, $\ket{\downarrow\uparrow}$ and $\ket{\downarrow\downarrow}$. We focus on the dynamics in the antiparallel subspace and define a pseudo-spin spanned by $\ket{\Uparrow}=\ket{\uparrow\downarrow}$ and $\ket{\Downarrow}=\ket{\downarrow\uparrow}$ (Methods) \cite{Zhao2011,Shi2013,Abobeih2018}. The pseudo-spin Hamiltonian is:
\begin{equation}
H = X\hat{I}_x + m_s Z \hat{I}_z, 
\end{equation}
in which $\hat{I}_z$ and $\hat{I}_x$ are spin–1/2 operators. $X$ is the dipolar coupling between the $^{13}$C spins, which creates the evolution $\ket{\Uparrow} \leftrightarrow \ket{\Downarrow}$ (flip-flops). $m_s = \{-1, 0, +1\}$ is the NV spin projection and $Z$ is the difference between the two NV-$^{13}$C hyperfine couplings (Methods).

Pair A and pair B are nearest-neighbour pairs oriented along the external magnetic field with $X_A = X_B = 2 \pi \cdot 2080.9900(3)$ Hz, $Z_A = 2 \pi \cdot 130(1)$ Hz and $Z_B = 2 \pi \cdot 91(2)$ Hz (see measurements below). Pair C has a larger spatial separation between the spins resulting in $X_C = 2 \pi \cdot 188.33(2)$ Hz, and $Z_C = 2 \pi \cdot 2802(2)$ Hz. In the following, we develop initialisation, control and measurement for pairs A and B, for which $X \gg Z$ (see Methods for pair C control, for which $Z \gg X$).

Previous work has demonstrated that the pseudo-spin of pairs can be detected through decoupling sequences that toggle the $m_sZ\hat{I}_z$ term by periodically inverting the NV electron spin (Fig. 1b) \cite{Zhao2011,Shi2013,Abobeih2018}. For an interpulse delay of $2\tau = \pi/\omega_r$, with $\omega_r = \sqrt{X^2+(Z/2)^2}$, the sequence is resonant with the pseudo-spin dynamics and the effective NV-pair interaction is of the form $\hat{S}_z\hat{I}_z$, with $\hat{S}_z$ the spin operator for the NV electron spin \cite{Zhao2011,Shi2013,Abobeih2018}. The NV center thus accumulates a phase that depends on the $z$-projection of the pseudo-spin. We use the NV center as a sensor to detect the spin pairs in its environment by sweeping $\tau$ (Fig. 1b) \cite{Zhao2011,Shi2013,Abobeih2018} and find the resonances for pair A and B ($\tau = 120.330$ \textmugreek s) and pair C ($\tau = 177.026$ \textmugreek s).

\begin{figure*}
\includegraphics[width = \textwidth]{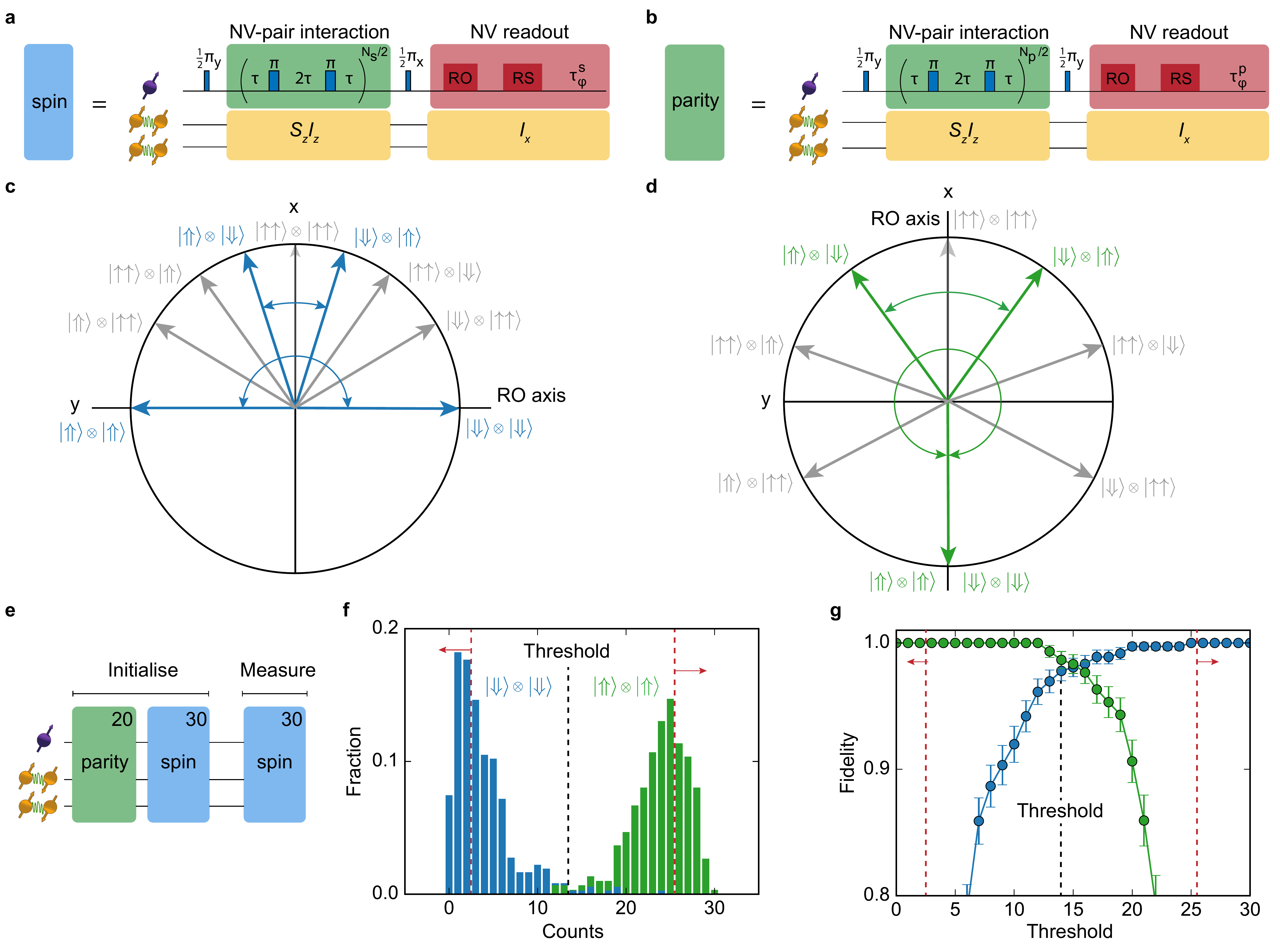}
\caption{\textbf{Projective spin and parity measurements for pairs A and B.} \textbf{a.} Sequence to measure the pseudo-spin states. The NV electron spin starts in $m_s = 0$. The  $\hat{S}_z\hat{I}_z$ interaction sequence ($\tau = 120.330$ \textmugreek s and $N_s = 14$) maps the state of the two pairs onto the NV spin. The NV spin is subsequently read out (RO) and reset (RS) to $m_s=0$. We synchronize subsequent repetitions by calibrating a waiting time $\tau_\phi^s = 323.5$ \textmugreek s to compensate for the $\hat{I}_x$ evolution during NV readout. This ensures that the full sequence duration is a multiple of $1/X$. \textbf{b.} Sequence to measure the pseudo-spin parity ($N_p = 20$, Supplementary Section VII). We set $\tau_\phi^p = 81$ \textmugreek s to synchronize subsequent measurements (sequence duration a multiple of $1/(2X)$). \textbf{c.} XY-plane of the NV Bloch sphere showing the possible phases accumulated in the spin measurement. The NV spin starts along $x$ and picks up a positive (negative) phase for a pair in $\ket{\Uparrow}$ ($\ket{\Downarrow}$) and no phase for a pair in a parallel state ($\ket{\uparrow\uparrow}$ or $\ket{\downarrow\downarrow}$). Reading out along the $y$-axis distinguishes the 4 pseudo-spin states (blue). Note that $\ket{\downarrow \downarrow}$ (not shown) gives the same result as $\ket{\uparrow \uparrow}$. \textbf{d.} XY-plane of the NV Bloch sphere under parity readout. The initial state ($x$-axis) and the readout axis ($x$-axis) are identical so that the parity of pair A and B is measured. \textbf{e.} Measurement sequence to calibrate single-shot readout and initialisation. The top right of each block indicates the number of repetitions. The optimal number of spin readouts is 30 (Supplementary Section VI). \textbf{f.} Conditional histograms for 30 spin readouts after initialisation in $\ket{\Uparrow}\ket{\Uparrow}$ (green) and $\ket{\Downarrow}\ket{\Downarrow}$ (blue). The initialisation conditions for the 30 preceding spin readouts are indicated in red. \textbf{g.} Combined initialisation and readout fidelity for $\ket{\Uparrow}\ket{\Uparrow}$ (green) and $\ket{\Downarrow}\ket{\Downarrow}$ (blue) for 30 spin readouts. We find an optimum of $F = 98.2(7) \%$ for a decision threshold of 14 out of 30.}
\label{Ext_sequence}
\end{figure*}

We start by developing projective single-shot measurements. Unlike all previous work, which was limited to manipulating mixed states of the parallel and antiparallel subspaces \cite{Abobeih2018}, these measurements enable us to initialise and measure the complete state of the spin pairs with high contrast. 

Our method is based on repeated non-destructive measurements and illustrated in Fig. 2. Each repetition comprises an interaction period between the NV and the pair pseudo-spin before optical readout. During the interaction the NV electron spin accumulates a positive (negative) phase if a pair is in $\ket{\Uparrow}$ ($\ket{\Downarrow}$). For a pair in the parallel subspace ($\ket{\uparrow \uparrow}$ or $\ket{\downarrow \downarrow}$), the NV spin does not accumulate any phase. We choose $\tau$ such that pairs A and B interact with the NV spin simultaneously. Therefore, the NV spin accumulates a phase that depends on which of the 16 possible states the two pairs are in (Fig. 2c). By repeatedly applying this sequence, we realize a projective measurement that can distinguish multiple states in a single shot and with high contrast.

We construct two types of measurements by setting different interaction times and NV readout axes (Fig. 2a,b). The `spin' measurement distinguishes the four pseudo-spin states ($\ket{\Uparrow\Uparrow}, \ket{\Uparrow\Downarrow}, \ket{\Downarrow\Uparrow}, \ket{\Downarrow\Downarrow}$; Fig. 2a,c). The `parity' measurement only distinguishes the pseudo-spin parity of the two pairs ($\{\ket{\Uparrow\Uparrow}, \ket{\Downarrow\Downarrow}\}$: even parity versus $\{\ket{\Uparrow\Downarrow}, \ket{\Downarrow\Uparrow} \}$: odd parity; Fig. 2b,d). Because the pseudo-spins evolve as $\ket{\Uparrow} \leftrightarrow \ket{\Downarrow}$ with a frequency $\sim X$ during the NV spin readout, each repetition must be timed to align the measurement axes. This synchronization of repeated non-destructive measurements to the system evolution is similar to the case of repeated measurements on individual spins, e.g. in the context of quantum algorithms \cite{Cramer2016, Liu2017}, atomic frequency locking and quantum Zeno dynamics \cite{Kalb2016}, and weak measurement sequences \cite{Cujia2019,Pfender2019}.

We combine these sequences to realize high-fidelity initialisation and measurement (Fig. 2e). We first apply the parity measurement sequence (20 repetitions) to herald preparation in an even parity state, and to exclude the cases for which one or both pairs are in their parallel subspace. Then, we apply a spin measurement (30 repetitions) to herald either $\ket{\Uparrow\Uparrow}$ or $\ket{\Downarrow\Downarrow}$. Finally, we measure the pseudo-spin state with another spin measurement (30 repetitions). The resulting conditional histograms show well-isolated distributions (Fig. 2f) and an optimization of the measurement decision threshold (Fig. 2g) yields a combined initialisation and readout fidelity of $98.2(7) \%$. We refer to the Supplementary Information for the full optimization procedure.

\begin{figure}
\includegraphics[width = \columnwidth]{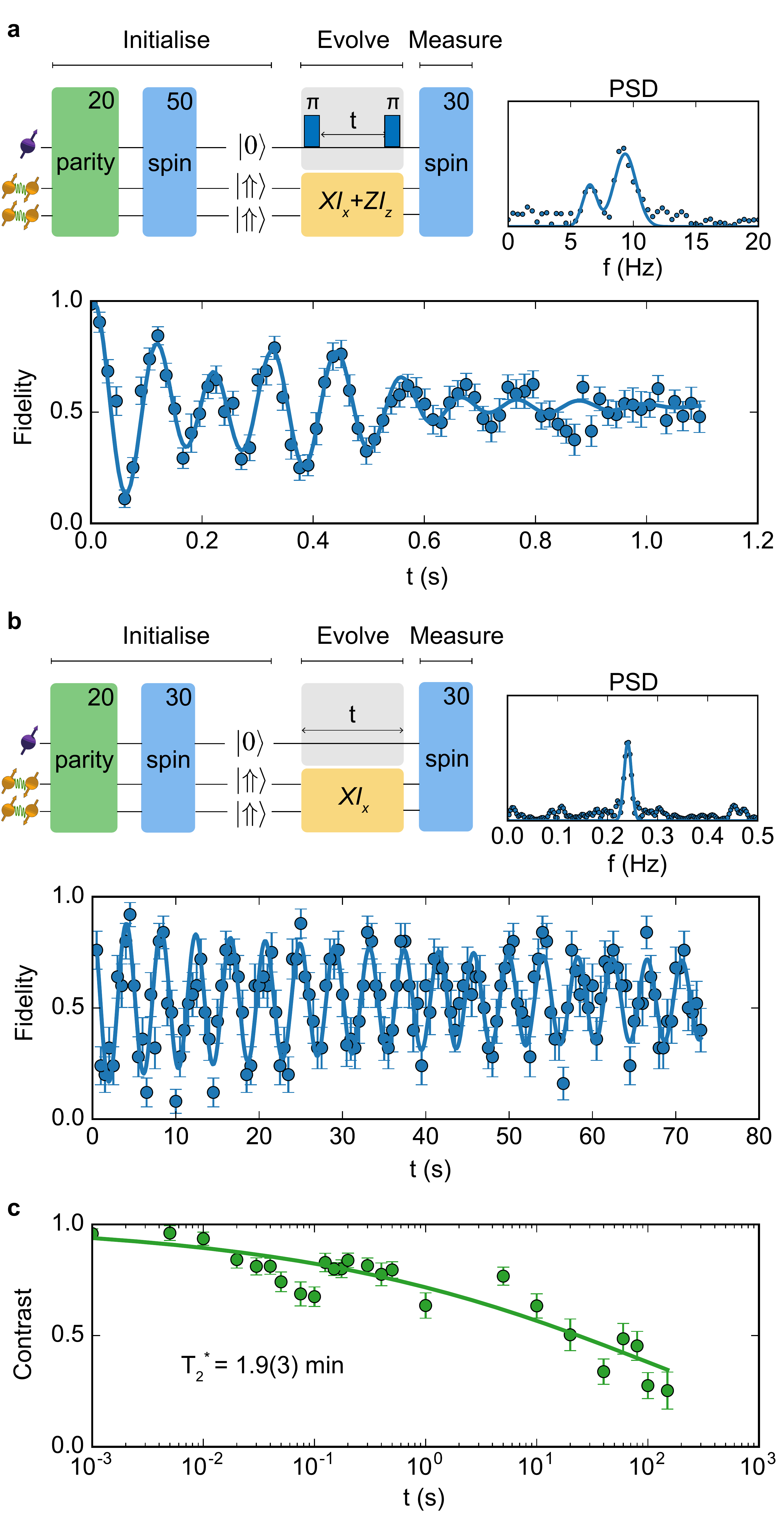}
\caption{\textbf{Coherence of pair A and B.} \textbf{a.} Ramsey measurement with the NV in $m_s = -1$. (Top left) Experimental sequence. (Top right) Fourier transform of the signal indicating two frequencies. From the data we obtain the coupling of the pairs to the NV: $Z_A = 2 \pi \cdot 130(1)$ Hz and $Z_B = 2 \pi \cdot 91(2)$ Hz. \textbf{b.} Ramsey measurement with the NV in $m_s = 0$. (Top left) Experimental sequence. (Top right) Fourier transform indicating a single frequency. From the data we obtain $X = 2 \pi \cdot 2080.9900(3)$ Hz. For a and b a detuning has been applied (Methods). \textbf{c.} Each data point corresponds to the amplitude of a Ramsey measurement in $m_s = 0$. A fit yields $T_2^* = 1.9(3)$ min, see Methods. The data deviates from a simple exponential decay, indicating that processes beyond pure dephasing contribute to decoherence (Supplementary Section I).}

\label{Ext_sequence}
\end{figure}

\begin{figure*}
\includegraphics[width = \textwidth]{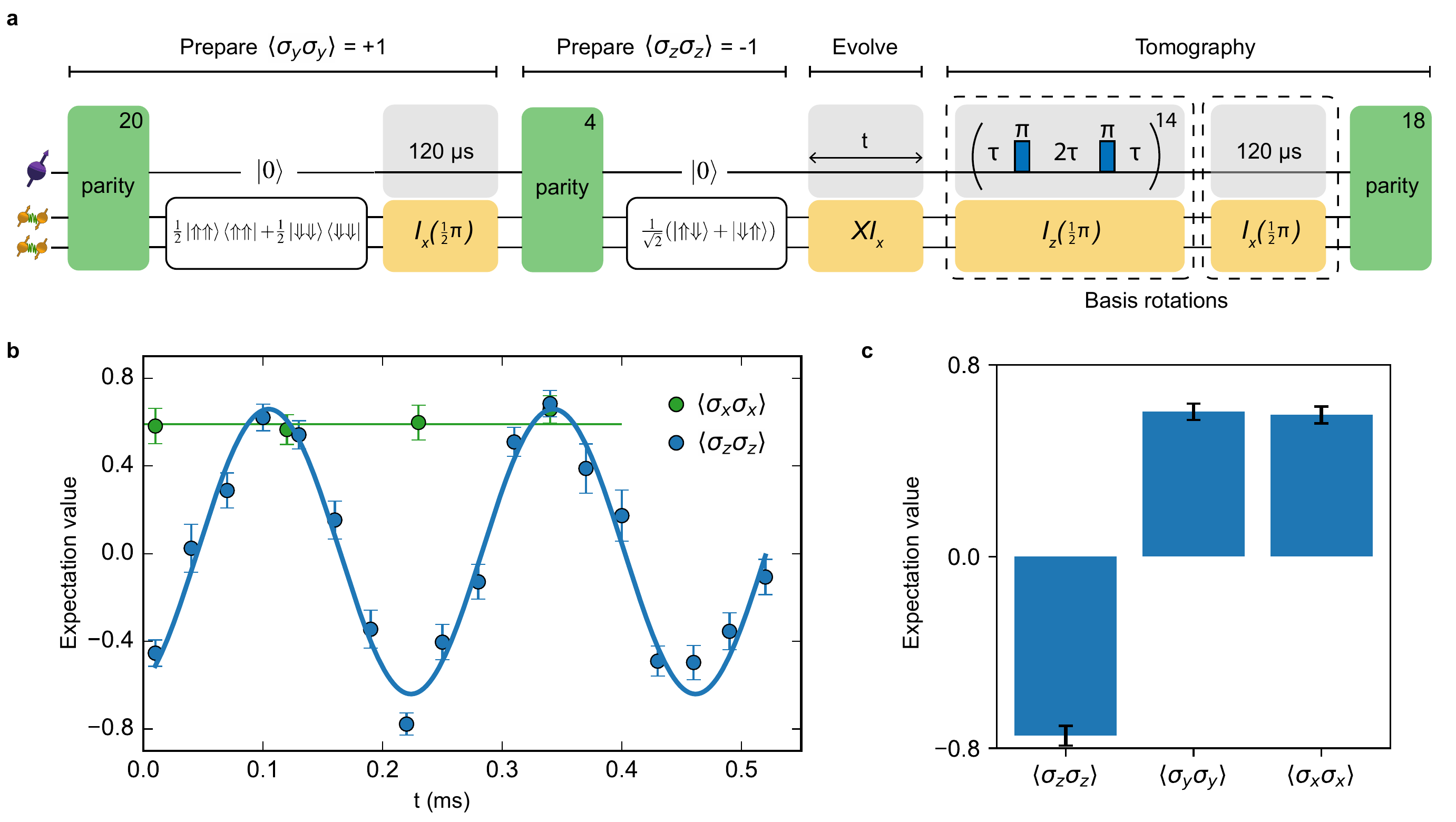}
\caption{\textbf{Entanglement of pair A and B.} \textbf{a.} Experimental sequence. We prepare the entangled state $\frac{1}{\sqrt{2}} (\ket{\Uparrow \Downarrow} + \ket{\Downarrow \Uparrow})$ by consecutively measuring $\langle \sigma_y \sigma_y \rangle$ and $\langle \sigma_z \sigma_z \rangle $. We herald on the +1 ($>14/20$ counts) and -1 ($<1/4$ counts) outcomes in the initialisation steps. Final operators are measured through optional basis rotations (dashed boxes) and a $\langle \sigma_z \sigma_z \rangle$ parity measurement. $I_{x/z}(\phi)$ stands for a rotation around $x/z$ by an angle $\phi$. \textbf{b.} Parity oscillations show a frequency of $4.20(4)$ kHz ($\sim 2X$). For $\langle \sigma_x \sigma_x \rangle$ no oscillation is observed as the pseudo-spin eigenstates are along $x$. \textbf{c.} Measurement of the three nonzero operators of the entangled state. The state fidelity is $F = (1 - \langle \sigma_z \sigma_z \rangle + \langle \sigma_y \sigma_y \rangle + \langle \sigma_x \sigma_x \rangle)/4 = 74(2)\%$. We use $t = 225$ \textmugreek s for measuring $\langle \sigma_z \sigma_z \rangle$ and $t = 105$ \textmugreek s for $\langle \sigma_y \sigma_y \rangle$. The results have not been corrected for readout infidelity.}
\label{Ext_sequence}
\end{figure*}

We now use the developed high-contrast measurements to investigate the coherence of pair A and B. First, we perform a free-evolution experiment with the NV spin in $m_s = -1$ (Fig. 3a), for which the NV-pair coupling is on. Because the pseudo-spin precession frequency $\sqrt{X^2+Z^2}$ is different for the pairs ($Z_A \neq Z_B$), this measurement reveals the presence of the two pairs and characterizes their couplings $Z$ to the NV. The two frequencies observed give $Z_A = 2 \pi \cdot 130(1)$ Hz and $Z_B = 2 \pi \cdot 91(2)$ Hz (Methods). We obtain the dephasing times from a Fourier transform (Fig. 3a): $T_{2,A}^* = 0.26(2)$ s and $T_{2,B}^* = 0.39(6)$ s (Methods). These values are one to two orders of magnitude larger than for individual $^{13}$C spins in the same sample \cite{Bradley2019}.

Second, we perform the same experiment with the NV spin in $m_s = 0$, so that the coupling to the NV is effectively turned off. Now both pairs precess with frequency $X_A = X_B = 2 \pi \cdot 2080.9900(3)$ Hz (Fig. 3b) and a coherent oscillation that extends past 70 s is observed. To extract the dephasing time, we measure the oscillation amplitude at various times (Fig. 3c). The resulting decay yields $T_2^*$ = 1.9(3) minutes, a four order-of-magnitude improvement over an individual spin \cite{Bradley2019} and the longest inhomogeneous dephasing time reported for any individually controllable quantum system \cite{Harty2014}.

We now elucidate the mechanisms which lead to these remarkable coherence properties. We add a magnetic-field noise term $\Delta Z(t)$ to the pseudo-spin Hamiltonian:
\begin{equation}
H = X\hat{I}_x + (m_s Z + \Delta Z(t)) \hat{I}_z.
\label{noise_hamiltonian}
\end{equation}
The first mechanism which enhances the coherence is the decoherence-free subspace \cite{Langer2005} formed by the pseudo-spin states. Because the spins are identical, $\Delta Z(t)$ is given by the fluctuations of the magnetic field \emph{difference} between the two spins. The atomic distance between the spins ensures near-complete immunity to noise from distant sources, such as the external magnetic field and the control signals. The main source of noise is the surrounding $^{13}$C spin bath. Hence, $\Delta Z (t)$ can be approximated as a Gaussian distribution with a correlation time $\tau_c$ \cite{AndersonP.W.;Weiss1953,Dobrovitski2009} and variance $b^2 = \frac{1}{4}\sum_k (A_k^{(1)} - A_k^{(2)})^2$, where $A_k^{(1)}$ ($A_k^{(2)}$) is the dipolar coupling of bath spin $k$ to the first (second) spin of the pair. We calculate the typical effective noise strength $b \sim 10$ Hz by numerically simulating many spin-bath configurations. This is a noise reduction by a factor $\sim 2$ due to the decoherence-free subspace (Supplementary Section II).

We first analyze the case of the NV electron spin in $m_s = -1$ (Fig. 3a), which enables us to extract the strength of the noise due to the spin bath. Because $X \gg Z \gg \Delta Z (t)$, the Hamiltonian can be approximated as (Supplementary Section I)
\begin{equation}
    H =  (\omega_{-1} + \frac{Z}{\omega_{-1}} \Delta Z (t)) \hat{I}_x,
    \label{noise_hamiltonian_approximation_ms1}
\end{equation}
with $\omega_{-1} = \sqrt{X^2+Z^2}$. Additionally, the NV spin creates a field gradient that suppresses spin flip-flops in the bath (a frozen core \cite{Bradley2019,Zhong2015}). Therefore, the noise can be treated as quasi-static and the signal decay is Gaussian \cite{AndersonP.W.;Weiss1953}, as experimentally observed (Fig. 3a). The dephasing time is given by \cite{AndersonP.W.;Weiss1953}
\begin{equation}
    T_2^* = \frac{\omega_{-1}}{Z}\frac{\sqrt{2}}{b}.
    \label{coherence_ms1}
\end{equation}
In this case, the coherence is enhanced by a factor $\frac{\omega_{-1}}{Z} \approx 20$ in addition to the enhancement by the  decoherence-free subspace. Finally, inserting the measured dephasing times into equation (\ref{coherence_ms1}) yields noise strengths $b_A = 2 \pi \cdot 13.9(2)$ Hz and $b_B = 2 \pi \cdot 12.5(4)$ Hz. These values are consistent with the inter-pair distance and $^{13}$C concentration (Supplementary Section II).

Second, we analyze the case with the NV electron spin in $m_s = 0$ (Fig. 3b). Because $X \gg \Delta Z (t)$, the Hamiltonian can be approximated as (Supplementary Section I) \cite{Dobrovitski2009}
\begin{equation}
H = X\hat{I}_x + \frac{\Delta Z^2(t)}{2X} \hat{I}_x.
\label{noise_hamiltonian_simplified}
\end{equation}
The eigenenergies are now first-order insensitive to $\Delta Z (t)$ as the spin pair forms a clock transition due to the coupling $X$, a second mechanism that enhances coherence. Note that the clock transition in this system does not require a specific magnetic field value, as the simultaneous decoherence-free subspace removes the dependence on global magnetic fields.

The decoherence-free subspace and clock transition alone cannot yet explain the observed $m_s = 0$ dephasing time. In particular, for quasi-static or slow noise the coherence would be limited to $\sim 10$ s (Supplementary Section I). However, the increased coherence, in combination with the lack of a frozen core for $m_s=0$, unlocks a new regime in which the nuclear-spin bath fluctuations become relatively fast ($\tau_c \ll X/b^2$). A mathematically equivalent Hamiltonian was analysed theoretically by Dobrovitski et al. \cite{Dobrovitski2009}. The resulting time constant is
\begin{equation}
    T_2^* = \frac{4X^2}{b^4 \tau_c}.\label{T2*_motional_narrowed}
\end{equation}
The dependence on the correlation time $\tau_c$ reveals a third  mechanism, similar to motional narrowing \cite{Dobrovitski2009}, that further enhances the coherence. Inserting the parameters obtained from the $m_s=-1$ measurement and a typical value for $\tau_c \sim 0.1$ s \cite{Cramer2016}, inhomogeneous dephasing times of $\sim 100$ s are predicted. Together these three mechanisms thus provide an explanation for the observed dephasing times.

To further analyse the different physical regimes that play a role, we investigate pair C (Fig. 1). Because $Z \gg X$, the dynamics are different and the clock transition can be switched on ($m_s = 0$) and off ($m_s = -1$) (Supplementary Section I). We develop complete control, initialisation and single-shot readout of such pairs in the Methods.

For evolution under $m_s = 0$, the situation is similar to pairs A and B. We find $T_2^* = 0.6(1)$ s, which is reduced compared to pairs A and B because the smaller coupling $X$ makes the clock transition less effective (Extended Data Fig. \ref{pair5_coherence}). Additionally, similar values obtained for spin echo ($T_2 = 0.7(1)$ s) and relaxation measurements ($T_1 = 0.9(2)$ s) indicate that relaxation plays a role in limiting the coherence (Supplementary Section I). For $m_s=-1$, a frozen core is formed and the clock transition is turned off, so that the noise $\Delta Z(t)$ affects the eigenfrequencies linearly. We find $T_2^* = 18(1)$ ms with Gaussian decay, indicating quasi-static noise \cite{AndersonP.W.;Weiss1953}, which is consistent with spin echo ($T_2 = 0.3(2)$ s $\gg T_2^*$) and relaxation measurements ($T_1 \gg 1$ s) (Extended Data Fig. \ref{pair5_coherence}). In this case, there is no significant coherence protection and the results are similar to individual $^{13}$C spins \cite{Bradley2019}.

Finally, we demonstrate the potential of the spin pairs as qubits by demonstrating an entangled state of pair A and pair B. We create entanglement through subsequent projective measurements of the $\sigma_y \sigma_y$  and $\sigma_z \sigma_z$ pseudo-spin parity (Fig. 4a). We herald on outcomes $\langle{\sigma_y \sigma_y}\rangle =+1$ and $\langle{\sigma_z \sigma_z}\rangle=-1$, so that the resulting state is $\frac{1}{\sqrt{2}}(\ket{\Uparrow \Downarrow} + \ket{\Downarrow \Uparrow})$. This state is a 4-spin entangled state $\frac{1}{\sqrt{2}}(\ket{\uparrow \downarrow \downarrow \uparrow}+\ket{\downarrow \uparrow \uparrow \downarrow})$ that encodes two qubits of information in two long-lived pseudo-spin states.

To characterize the resulting state we first measure parity oscillations by varying the evolution time $t$ (Fig. 4a). The observed frequency is 4.20(4) kHz, which equals $2X/2\pi$, as expected (Fig. 4b). To determine the state fidelity, we measure the pseudo-spin parity operators $\langle \sigma_x \sigma_x \rangle$, $\langle \sigma_y \sigma_y \rangle$ and $\langle \sigma_z \sigma_z \rangle$. We realize the required single-qubit rotations through waiting times (for $x$-rotations) and dynamical decoupling sequences with the NV spin in an eigenstate (for $z$-rotations) (Fig. 4a). Figure 4c shows the resulting expectation values, which yield a fidelity $F = 0.74(2)$, confirming entanglement ($F > 0.5$). This result highlights the high-fidelity initialisation, control, and non-destructive measurements realized.

In conclusion, we have developed complete control over multiple nuclear-spin pairs. These spin pairs provide inherently long-lived quantum states due to a unique combination of three physical phenomena: a decoherence-free subspace, a clock transition and a variant of motional narrowing. This inherent coherence protection makes spin pairs promising qubits for a variety of applications, including quantum sensing \cite{Degen2017} and quantum information processing \cite{Bradley2019}. In particular, due to the small effective coupling to the NV spin, they might provide robust quantum memories for optically connected quantum networks \cite{Hensen2015, Reiserer2016}. Such spin pairs are available for most NV centers  (Supplementary Section IV) and are present in a variety of other materials. Therefore, our results reveal a new, promising, and abundantly available resource for quantum science and technology.\\

\noindent\textbf{METHODS}

\noindent\textbf{Sample and NV center.} The experiments are performed on a naturally occurring NV center in a cryogenic confocal microscope (3.7 K). The diamond was homoepitaxially grown using chemical vapor deposition and cleaved along the $\langle 111 \rangle$ axis (Element Six). There is a natural abundance ($1.1 \%$) of $^{13}$C. The NV centre was selected on the absence of strongly coupled $^{13}$C spins exceeding $\sim 500$ kHz hyperfine coupling, but without any other criteria on the spin environment or spin pairs.\\
The NV electron spin has a dephasing time of $T_2^* = 4.9(2)$ \textmugreek s and a spin echo time of $T_2 = 1.182(5)$ ms. The electron relaxation ($T_1 > 1$ h) at this temperature is negligible \cite{Abobeih2018}. We measure the NV spin state in a single shot using spin-selective optical readout \cite{Cramer2016}. The readout fidelities are 0.905(2) (0.986(2)) for the $m_s = 0$ ($m_s = -1$) state with an average fidelity of $F = 0.946(1)$. The dynamical decoupling sequences follow the XY8-scheme to mitigate pulse errors \cite{Gullion1990}.

\noindent\textbf{Pseudo-spin Hamiltonian.} The Hamiltonian for two $^{13}$C spins in the vicinity of an NV center in an appropriate rotating frame and under the secular approximation can be written as: 
\begin{align}
\begin{split}
    H &= \omega_L I_z^{(1)} + \omega_L I_z^{(2)} + m_s\mathbf{A}^{(1)} \cdot \mathbf{I}^{(1)}\\
    &+ m_s\mathbf{A}^{(2)} \cdot \mathbf{I}^{(2)} + H_D.
\end{split}
\end{align}
where $\omega_L = \gamma_c B$ is the $^{13}$C spin Larmor frequency, with $\gamma_c$ the $^{13}$C gyromagnetic ratio. $B$ is the magnetic field along the NV-axis. $\mathbf{I}^{(i)}$ are the spin-$\frac{1}{2}$ operators acting on spin $i$, $m_s = \{-1,0,+1\}$ are the NV electron spin states and $\mathbf{A}^{(i)} = [A_x, A_y, A_z]$ is the NV-$^{13}$C hyperfine interaction vector of spin $i$. $H_D$ is the dipolar interaction between two $^{13}$C spins. For a large magnetic field compared to the dipolar ($X$) and hyperfine couplings ($A^{(1)},A^{(2)}$) $H_D$ can be written as:
\begin{align}
    H_D &= X (3 I_z^{(1)} I_z^{(2)} - \mathbf{I}^{(1)} \cdot \mathbf{I}^{(2)})\\
    X &= \frac{\mu_0 \gamma_c \gamma_c \hbar}{8 \pi \abs{\mathbf{r}_{12}}^3}(1-3 \cos^2 \theta_{12}),
\end{align}
where $\mu_0$ is the vacuum permeability, ${\mathbf{r}_{12}}$ is the vector between the two $^{13}C$ atoms and $\theta_{12}$ the angle between the magnetic field axis and the pair axis. Since $\omega_L = \gamma_c B = 2 \pi \cdot 432.140$ kHz is large compared to the dipolar ($X$) and hyperfine couplings ($A^{(1)},A^{(2)}$), the antiparallel states $\ket{\uparrow\downarrow}$ and $\ket{\downarrow\uparrow}$ form an isolated subspace in which we define a pseudo-spin $\frac{1}{2}$ as $\ket{\Uparrow}=\ket{\uparrow\downarrow}$ and $\ket{\Downarrow}=\ket{\downarrow\uparrow}$ \cite{Zhao2011,Zhao2012,Shi2013,Abobeih2018}. The Hamiltonian in this subspace is given by \cite{Zhao2011,Zhao2012}
\begin{equation}
    H = X\hat{I}_x + m_s Z\hat{I}_z.
\end{equation}
$Z$ originates from the difference of the hyperfine couplings of the two spins to the NV electron spin, and is given by \cite{Zhao2012}
\begin{equation}
    Z = Z_\parallel + Z_\perp = A^{(1)}_\parallel - A^{(2)}_\parallel + \frac{(A^{(1)}_\perp)^2 - (A^{(2)}_\perp)^2}{\gamma_c B},
\end{equation}
where $A^{(i)}_\parallel = A_z^{(i)}$ and $A^{(i)}_\perp = \sqrt{(A_x^{(i)})^2 + (A_y^{(i)})^2}$ for spin $i$ of the pair.

\noindent\textbf{Coherence of pair A and B.} The Ramsey data in Fig. 3a ($m_s=-1$) is fitted to
\begin{align}
\begin{split}
   F(t) &= a + \exp(-(t/T)^n)[A\cos(2 \pi f_A t + \phi_A)\\
   &+ B\cos(2 \pi f_B t + \phi_B)].
\end{split}
\end{align}
We obtain $T = 0.53(4)$ s, $n = 2.1(4)$, $f_A = 9.07(6)$ Hz and $f_B = 7.0(1)$ Hz (measured with a 10 Hz detuning with respect to 2086 Hz). Using $f = \sqrt{X^2+Z^2}$ and $X = 2 \pi \cdot 2080.9900(3)$ Hz, the values for $f_A$ and $f_B$ yield $Z_A = 2 \pi \cdot 130(1)$ Hz and $Z_B = 2 \pi \cdot 91(2)$ Hz. The observed shape of the decay ($n = 2.1(4)$) is in agreement with the predicted Gaussian ($n=2$) decay for quasi-static noise (Supplementary Section I). 

To extract the dephasing times we fit the Fourier transform in Fig. 3a to 
\begin{align}
\begin{split}
    F(f) = a &+ A \exp(-(f+f_A)^2/2\sigma_A^2)\\ 
    &+ B \exp(-(f+f_B)^2/2\sigma_B^2).
\end{split}
\end{align}
We find $\sigma_A = 0.88(6)$ Hz and $\sigma_B = 0.57(9)$ Hz which gives $T_{2,A}^* = 1/(\sqrt{2}\pi\sigma_A) =  0.26(2)$ s and $T_{2,B}^* = 0.39(6)$ s.

The Ramsey data in Fig. 3b ($m_s=0$) is fitted to
\begin{equation}
    F(t) = \exp(-(t/T)^n)\cos(2 \pi f t + \phi).
\end{equation}
We obtain $T = 98(44)$ s, $n = 0.5(4)$ and $f = 0.2400(3)$ Hz (measured with a 0.25 Hz detuning with respect to 2081 Hz). Therefore we obtain $X = 2 \pi \cdot 2080.9900(3)$ Hz. Note that the precise value obtained for $X$ deviates from simple theoretical estimates and is analyzed in Supplementary Section X. The Fourier transform is fitted to 
\begin{equation}
    F(f) = a + A \exp(-(f+f_0)^2/2\sigma_0^2).
\end{equation}
We obtain $f_0 = 0.2402(3)$ Hz and $\sigma_0 = 0.0074(3)$ Hz.

The data in Fig. 3c is fitted to $\exp(-(t/T)^n)$ obtaining $T = 1.9(3)$ min and $n = 0.23(2)$. Note that $n$ deviates from the simple exponential decay ($n=1$) associated to equation \ref{T2*_motional_narrowed}, indicating that other effects beyond pure dephasing contribute to the decoherence (Supplementary Section I).

\noindent \textbf{Pseudo-spin control methods.} The control methods for pairs A and B are given in the main text and summarized in Extended Data Fig. \ref{control_schematics}.   

Pair C has a dipolar coupling $X = 2 \pi \cdot 188.33(2)$ Hz and a hyperfine difference $Z = 2 \pi \cdot 2802(2)$ Hz. Therefore, $Z \gg X$, in contrast to pairs A and B for which $X \gg Z$. This changes the dynamics in two ways. First, for $m_s =-1$, $Z$ is the dominant term in the pair frequency $\omega_{-1} = \sqrt{X^2+Z^2}$ and thus sets the location of the resonance in Fig. 1b. Second, the effective NV-pair interaction during the dynamical decoupling sequence becomes $\hat{S}_z\hat{I}_x$  \cite{Abobeih2018} (Extended Data Fig. \ref{control_schematics}).

We implement two types of projective measurements on pair C (see Extended Data Fig. \ref{pairC_control}). The spin measurement sequence distinguishes the pseudo-spin states $\frac{1}{\sqrt{2}}(\ket{\Uparrow} \pm \ket{\Downarrow})$. The parity measurement sequence distinguishes between the parallel ($\ket{\uparrow \uparrow}$,$\ket{\downarrow \downarrow}$) and antiparallel ($\ket{\Uparrow }$,$\ket{\Downarrow}$) subspaces of the pair. We obtain high-fidelity initialisation and readout by repeatedly applying these sequences (Supplementary Section VIII, IX).

For spin pairs with $Z \gg X$, the timing of repetitions is complicated by the fact that the $m_s = 0$ and $m_s = -1$ evolution frequencies and eigenstates differ significantly. Here, we mitigate this by minimizing the NV readout time (RO, $\sim 5$ \textmugreek s) and applying a fast reset of the NV spin (RS), so that the potential time spent in $m_s = -1$ is small. Because the states that the measurement projects onto ($\frac{1}{\sqrt{2}}(\ket{\Uparrow} \pm \ket{\Downarrow})$) are eigenstates of the $m_s = 0$ evolution, there is no timing requirement after resetting the NV and we simply concatenate subsequent measurements.

For pairs A and B ($X \gg Z$), we use free evolution and dynamical decoupling sequences to realize universal single-qubit control for the pseudo-spins (Fig. 4; Extended Data Fig. \ref{control_schematics}). For pair C ($Z \gg X$) all single-qubit operations can be obtained by letting the system evolve freely. Evolution with the NV electron spin in $m_s = 0$ implements a rotation around the $x$-axis, and evolution under $m_s = -1$ realizes a rotation around the $z$-axis (Extended Data Fig. \ref{control_schematics}). Note that the $z$-axis rotation is approximate as $Z$ is finite. In principle, this can be corrected for but this is not done here. We use the pair C control to measure the pseudo-spin dephasing time $T_2^*$, the spin echo time $T_2$ and the relaxation time $T_1$ in both NV electron spin states, see Extended Data Fig. \ref{pair5_coherence}.

\noindent \textbf{Spectroscopy and control of the full Hilbert space.}
Most of the work presented is focused on initialising, controlling and measuring the states in the antiparallel subspace of spin pairs, i.e. $\ket{\uparrow \downarrow}$ and $\ket{\downarrow \uparrow}$. In Extended Data Fig. \ref{rf}, we demonstrate that the entire Hilbert space of the spin pairs can be controlled by RF driving the single-spin-flip transitions of pair C.

The single-spin transition frequencies are $\omega_1 = 2 \pi \cdot 429.314(5)$ kHz and $\omega_2 = 2 \pi \cdot 432.122(7)$ kHz (Extended Data Fig. \ref{rf}a). Since the frequency of a single-spin transition in $m_s = -1$ is $\omega \approx \omega_L - A_\parallel$, this yields $A^{(1)}_\parallel = 2 \pi \cdot 2826(5)$ Hz and $A^{(2)}_\parallel = 2 \pi \cdot 18(7)$ Hz. Note that these values assume that $A_\perp$ is of similar magnitude, so that it can be neglected. The frequencies observed are consistent with the characteristic $^{13}$C frequencies ($\omega_L = 2 \pi \cdot 432.140$ kHz), further corroborating our assignment of $^{13}$C-$^{13}$C pairs as the source of the signals. 

These results also demonstrate selective initialisation, control and measurement of an individual carbon spin with negligible coupling to the NV by using its coupling to neighbouring spins. Spin 2 couples negligibly to the NV ($18(7)$ Hz), so that it overlaps in precession frequency with most of the spin bath. Nevertheless, it can be initialised and controlled selectively by using the NV to directly detect its flip-flops with spin 1 (i.e. pseudo-spin dynamics).

\noindent\textbf{Acknowledgements.}
We thank V.V. Dobrovitski for valuable discussions. This work was supported by the Netherlands Organisation for Scientific Research (NWO/OCW) through a Vidi grant and as part of the Frontiers of Nanoscience (NanoFront) programme. This project has received funding from the European Research Council (ERC) under the European Union’s Horizon 2020 research and innovation programme (grant agreement No. 852410). This project (QIA) has received funding from the European Union’s Horizon 2020 research and innovation programme under grant agreement No 820445. B.P. acknowledges financial support through a Horizon 2020 Marie Skłodowska-Curie Actions global fellowship (COHESiV, Project Number: 840968) from the European Commission.

\noindent\textbf{Author contributions.} 
HPB and THT devised the experiments. HPB and MHA performed the experiments. HPB, MHA, BP, MJD, SJHL and THT analyzed the data. MHA, CEB, JR and HPB prepared the experimental apparatus. MM and DJT grew the diamond sample. HPB and THT wrote the manuscript with input from all authors. THT supervised the project.

\noindent{\textbf{Competing Interests}. The authors declare no competing interests.}

\noindent\textbf{Data availability}. The data that support the findings of this study are available from the corresponding author upon request.

\bibliographystyle{naturemag}
\bibliography{bib3.bib}

\renewcommand{\figurename}{\textbf{Extended Data Fig.}}
\setcounter{figure}{0}

\begin{figure*}[h]
\includegraphics[width = \textwidth]{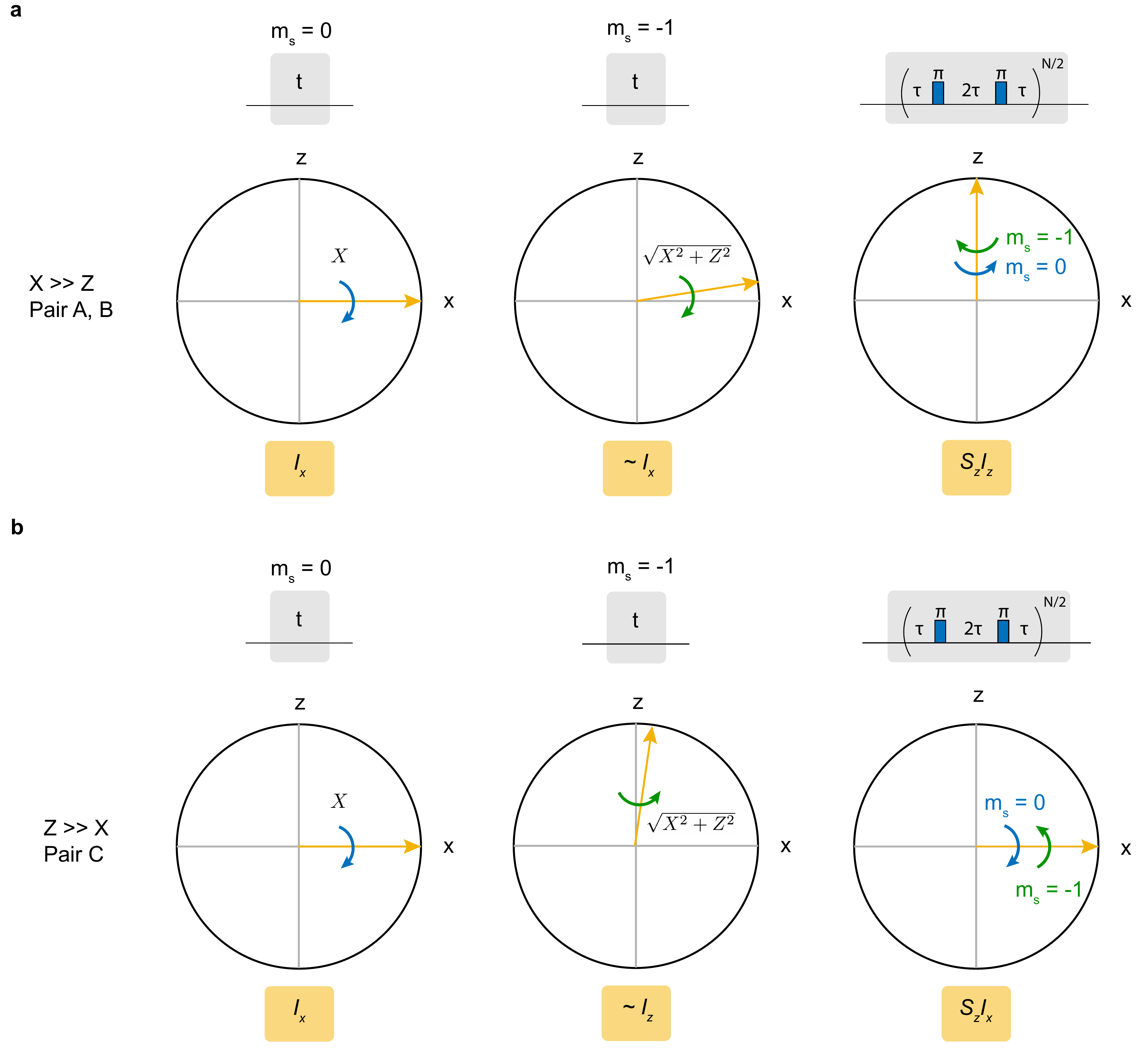}
\caption{\textbf{Overview of pair pseudo-spin dynamics during various sequences on the NV electron spin.} \textbf{a.} Pseudo-spin dynamics of pairs with $X \gg Z$ (pair A, B). The top row indicates the sequence performed on the NV electron spin, the middle row the corresponding pair dynamics in the XZ-plane of the pseudo-spin Bloch sphere and the bottom row indicates the effective pseudo-spin Hamiltonian term under that sequence. For the left two columns the rotation frequencies are given inside the Bloch spheres. From left to right the sequences are free evolution in $m_s = 0$ and $m_s = -1$, and a dynamical decoupling sequence with $\tau$ resonant, i.e. $2\tau = \pi/ \sqrt{X^2+(Z/2)^2}$. Rotations that are unconditional on the NV electron spin state can be obtained by setting $\tau = \pi/ \sqrt{X^2+(Z/2)^2}$ (unconditional $z$-rotation) and by setting $\tau$ far off-resonant (unconditional $x$-rotation), but these are not shown or used here. Note that the $z$-rotation frequency depends on the hyperfine field difference $Z$, so that pair A and B can, in principle, be controlled individually. \textbf{b.} Pseudo-spin dynamics of pairs with $Z \gg X$ (pair C). Like above, $z$- and $x$-rotations that are unconditional on the NV electron-spin state can be obtained by setting different values for $\tau$ (not shown). Together these operations enable universal control of the system consisting of the three pseudo-spins and the NV center.}
\label{control_schematics}
\end{figure*}

\begin{figure*}
\includegraphics[scale=0.35]{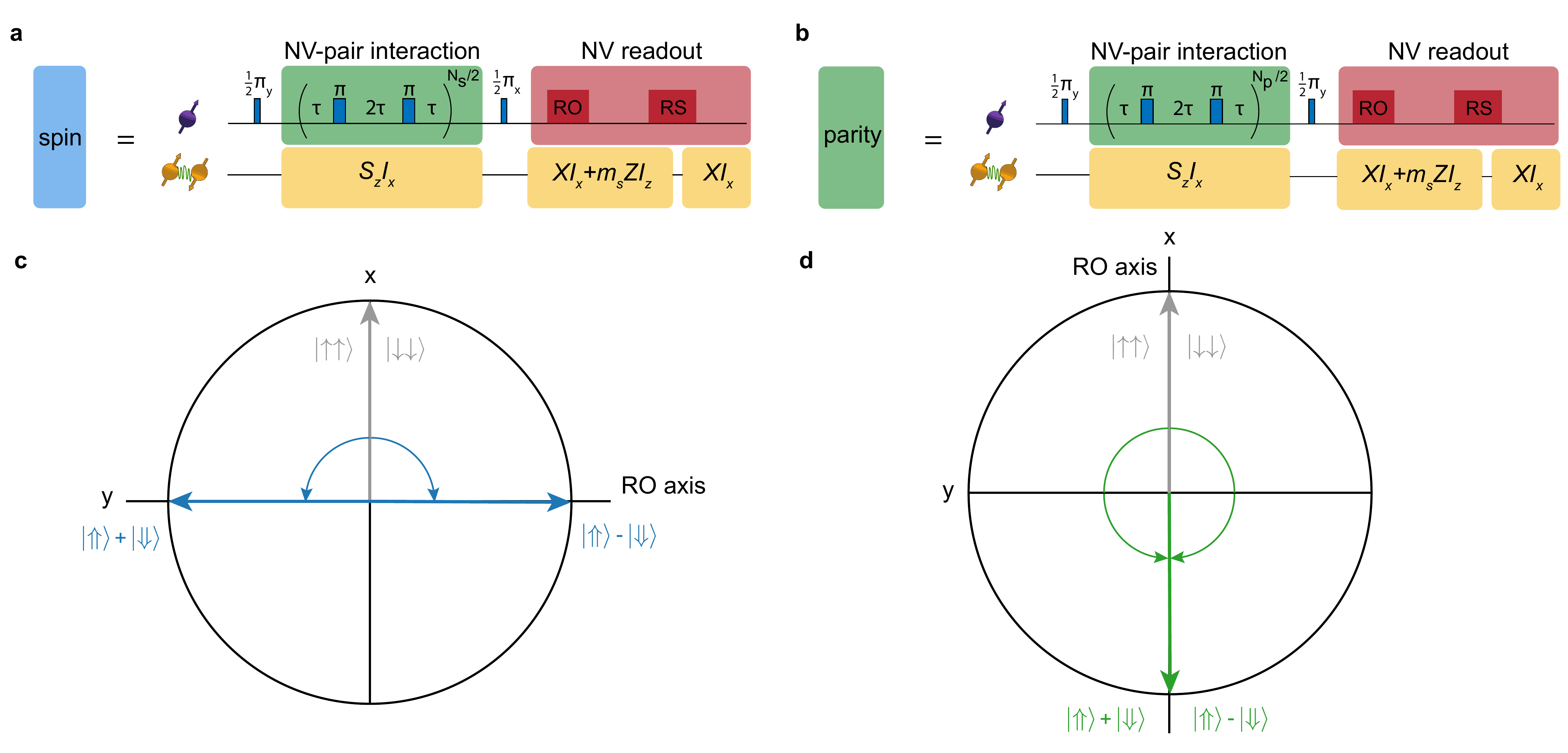}
\caption{\textbf{Projective spin and parity measurements for Pair C.} \textbf{a.} Sequence to measure the spin state of pair C. The NV starts in $m_s = 0$. Because $Z\gg X$, the effective interaction between the NV spin and pseudo-spin is $\hat{S_z}\hat{I_x}$. During the NV readout (RO) and reset (RS), the NV can spend an unknown time in $m_s = -1$ which causes dephasing of the pair spin. Additionally pair C undergoes a deterministic $z$-rotation for any known time spent in $m_s = -1$. To minimize these effects, we use a fast readout and reset. $N_s = 8$. \textbf{b.} Sequence to measure the parity of the two spins that make up pair C. Note that in this case the timing of the sequence is unimportant, as evolution in $m_s=-1$ or $m_s=0$ does not change the parity. $N_p = 14$. \textbf{c.} XY-plane of the NV Bloch sphere during the NV-pair interaction in a. The NV picks up a positive or negative phase depending on the $x$-projection of the pair pseudo-spin and no phase when the pair is in the parallel subspace. \textbf{d.} XY-plane of the NV Bloch sphere during the NV-pair interaction in b.}
\label{pairC_control}
\end{figure*}

\begin{figure*}
\includegraphics[scale=0.3]{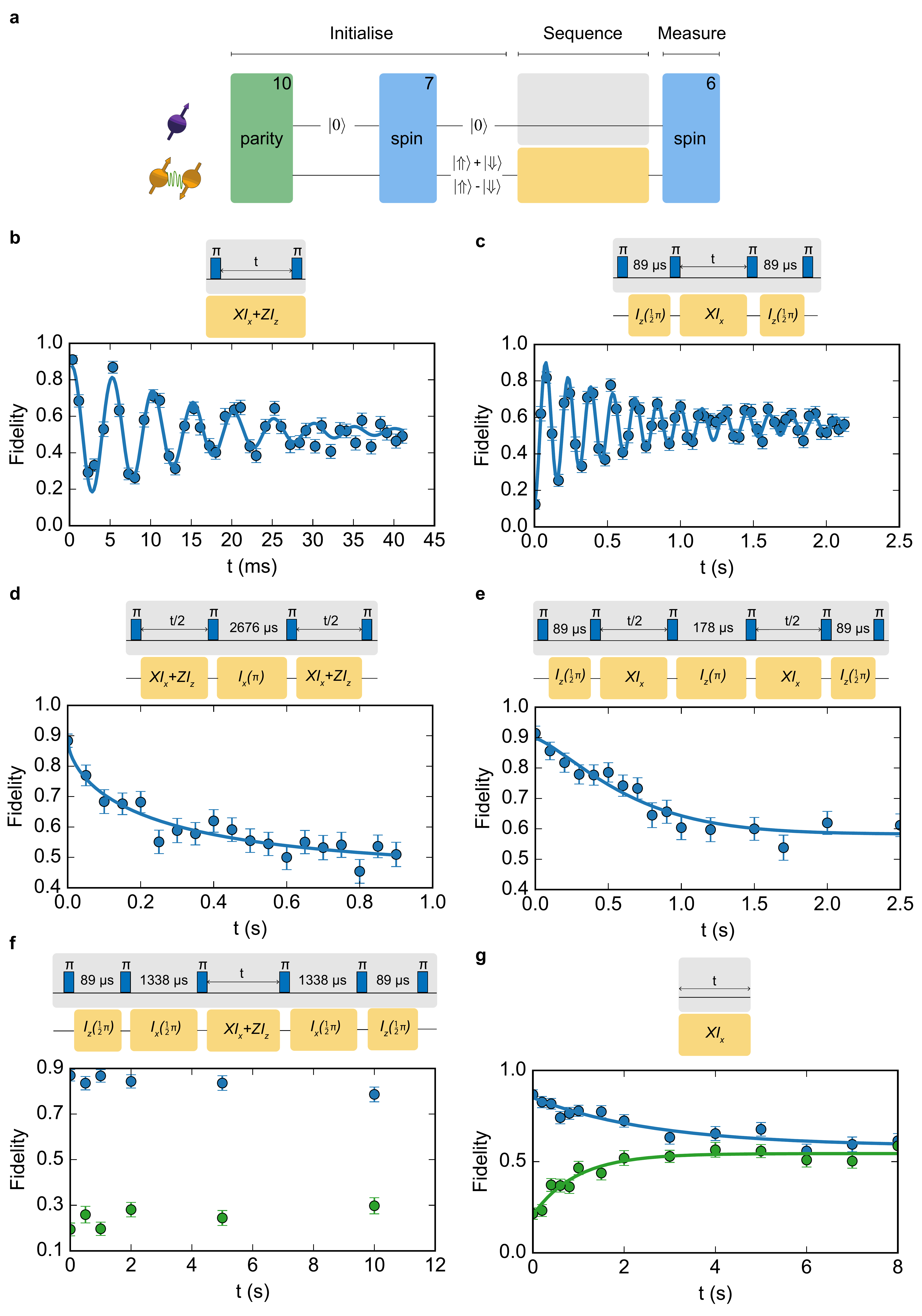}
\caption{\textbf{Coherence and relaxation of pair C.} \textbf{a.} Measurement sequence. First, we use 10 parity readouts to herald the pair in the antiparallel subspace (condition $>7/10$). Then, we use 7 spin readouts to initialise the pair in  $\frac{1}{\sqrt{2}}(\ket{\Uparrow} + \ket{\Downarrow})$ ($>4/7$, blue data) or $\frac{1}{\sqrt{2}}(\ket{\Uparrow} - \ket{\Downarrow})$ ($<3/7$, green data). The various evolution sequences are given as insets in panels b-g. Finally, 6 spin readouts are used to readout the spin state. Panels b,c are fitted to $ F(t) = a + A e^{-(t/T)^n} \cos (2 \pi f t + \phi)$ and panels d,e,g to $F(t) = a + A e^{-(t/T)^n}$. \textbf{b.} Ramsey measurement in $m_s = -1$. $T_2^* = 0.018(1)$ s, $n = 1.4(2)$ and $f = 2808(1)$ Hz (measured with a 200 Hz detuning with respect to 2807 Hz). \textbf{c.} Ramsey measurement in $m_s = 0$. $T_2^* = 0.6(1)$ s, $n = 0.7(1)$ and $f = 188.33(2)$ Hz (measured with a 5 Hz detuning with respect to 186.8 Hz). \textbf{d.} Spin echo measurement in $m_s = -1$. $T_2 = 0.3(2)$ s and $n = 0.6(2)$. \textbf{e.} Spin echo measurement in $m_s = 0$. $T_2 = 0.7(1)$ s and $n = 1.3(3)$. \textbf{f.} Relaxation measurement in $m_s = -1$. $T_1 \gg 1$ s. \textbf{g.} Relaxation measurement in $m_s = 0$. $T_1 = 3.6(7)$ s and $n = 0.8(2)$ for the blue data ($\frac{1}{\sqrt{2}}(\ket{\Uparrow} + \ket{\Downarrow})$). $T_1 = 0.9(2)$ s and $n = 1.0(2)$ for the green data ($\frac{1}{\sqrt{2}}(\ket{\Uparrow} - \ket{\Downarrow})$). The relaxation times are different for the two eigenstates, indicating a mechanism that depends on whether the state is a singlet or triplet.}
\label{pair5_coherence}
\end{figure*}

\begin{figure*}
\includegraphics[scale=0.6]{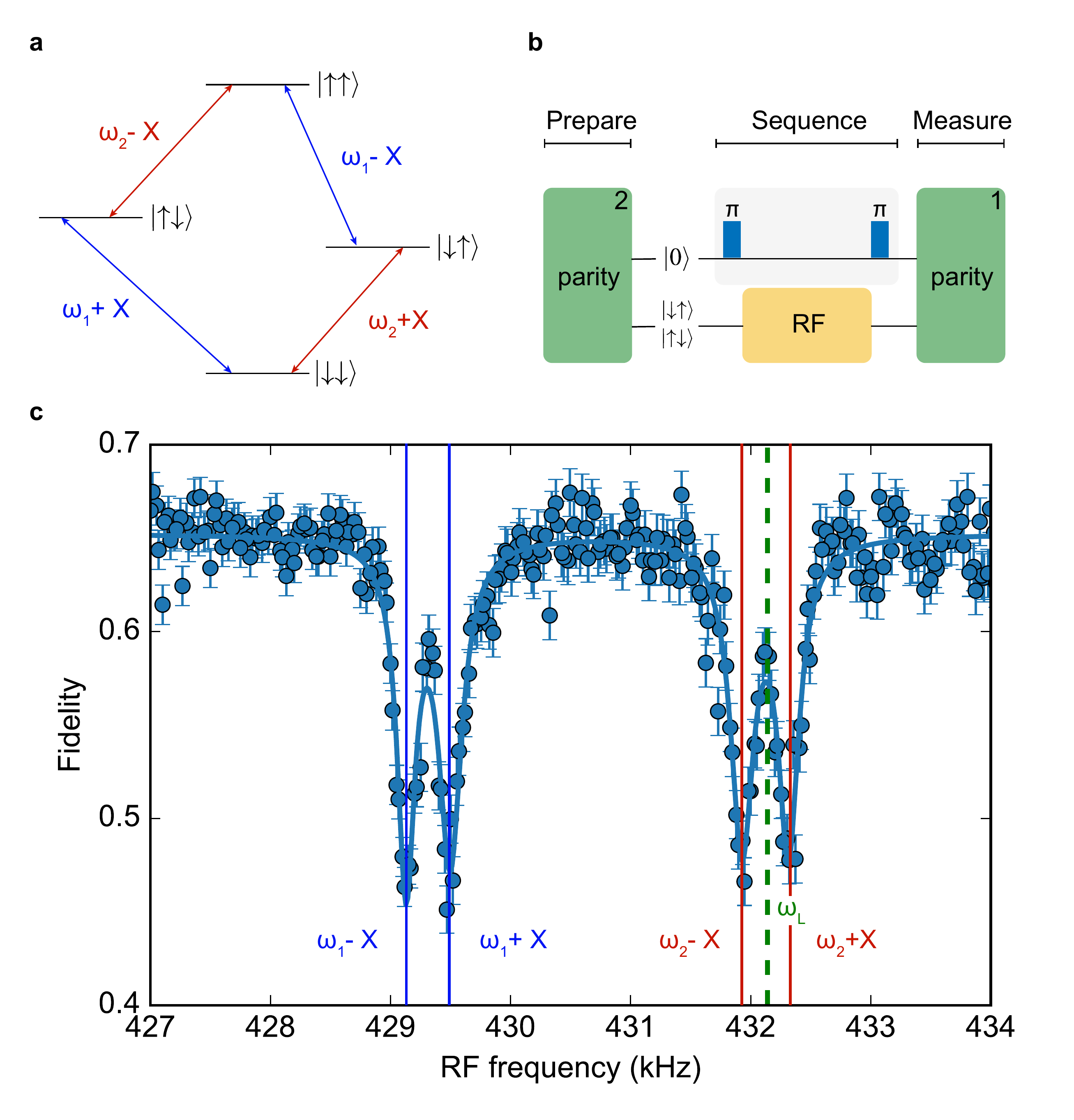}
\caption{\textbf{Spectroscopy and control of the complete pair Hilbert space.} \textbf{a.} Level diagram for pair C with the electron spin in $m_s = -1$. $\omega_1$ ($\omega_2$) is the frequency associated to the first (second) spin of the pair with the NV in $m_s = -1$. \textbf{b.} Sequence to reveal the transitions between the subspaces. First, the pair is initialised in the antiparallel subspace through a parity measurement, then an RF pulse with variable frequency with the NV in $m_s = -1$ is applied, and finally the subspace population is measured using another parity measurement. If the frequency of the RF pulse is resonant with a single-spin transition, the spin pair changes its subspace. \textbf{c.} Measurement result. Four transitions are observed corresponding to the marked transitions in a. The green dashed line corresponds to the bare Larmor frequency $\omega_L = 2 \pi \cdot 432.140$ kHz. We fit the data to four Lorentzians and extract $\omega_1 = 2 \pi \cdot 429.314(5)$ kHz, $\omega_2 = 2 \pi \cdot 432.122(7)$ kHz. For the left (right) dips we also obtain $X = 2 \pi \cdot 184(3)$ $(2 \pi \cdot 194(4))$ Hz. These results corroborate the assignment of the signals to $^{13}$C pairs and enable complete control over the full pair state.}
\label{rf}
\end{figure*}

\end{document}

\title{Supplementary information for \\ ``Coherence and entanglement of inherently long-lived spin pairs in diamond''}

\date{\today}
\maketitle
\tableofcontents

\newpage

\section{Spin-pair coherence regimes} \label{coherence_regimes}

In this section we provide the derivations for the spin-pair coherence alongside a more detailed discussion. The key idea is that different physical regimes are accessed depending on the NV electron-spin state and the spin-pair parameters. In particular, the analysis shows that the long coherence times observed for the $m_s=0$ state are made possible by a unique combination of a decoherence-free subspace, a clock transition and a variant of motional narrowing.   

We first focus on interactions and noise that cause dephasing or relaxation within the pseudo-spin subspace. Potential relaxation out of the pseudo-spin subspace, for example due to flip-flops with the surrounding bath spins, is discussed separately below. The Hamiltonian for a pair in the pseudo-spin subspace (spanned by $\ket{\Uparrow}=\ket{\uparrow\downarrow}$ and $\ket{\Downarrow}=\ket{\downarrow\uparrow}$), including a magnetic-field noise term $\Delta Z (t)$, is
\begin{equation}
    H = X\hat{I}_x + (m_s Z + \Delta Z(t))\hat{I}_z,
\end{equation}
where $X$ is the dipolar coupling between the two $^{13}$C spins, $m_s = \{0,-1\}$ the electron state ($m_s = +1$ is not populated during the experiments) and $Z$ is due to the hyperfine field gradient induced by the electron spin. Because the two spins are identical, the pseudo-spin forms a decoherence-free subspace. It is therefore only sensitive to field gradients. Noise from distant sources, such as fluctuations of the applied magnetic field (distance to the closest magnet: $\sim$ 1 cm) and noise from the control electronics on the on-chip stripline (distance: $\sim 10\ \mu$m) has negligible influence on the pseudo-spin dynamics, see section \ref{effect_external_noise}. Therefore, the surrounding $^{13}$C bath is the main source of noise and the origin of $\Delta Z (t)$.

The interaction between two central spins and a bath can be described as
\begin{equation}
    H_{int} = I_z^{(1)} \sum_{k} A^{(1)}_k I_z^{(k)} + I_z^{(2)} \sum_{k} A^{(2)}_k I_z^{(k)} + H_B,
\end{equation}
where $A^{(1)}_k$ ($A^{(2)}_k$) is the dipolar coupling of bath spin $k$ to spin 1 (2) of the pair and $I_z^{(i)}$ the nuclear spin-1/2 $z$-operator on spin $i$. $H_B$ describes the intrabath coupling.  We rewrite this interaction as
\begin{align*}
    H_{int} &= \frac{1}{2} \Big(\sum_k A^{(1)}_k + \sum_k A^{(2)}_k\Big) \Big(I_z^{(1)} + I_z^{(2)}\Big) I_z^{(k)} + \frac{1}{2} \Big(\sum_k A^{(1)}_k - \sum_k A^{(2)}_k\Big) \Big(I_z^{(1)} - I_z^{(2)}\Big) I_z^{(k)} + H_B.
\end{align*}
Only the latter term $\Big(I_z^{(1)} - I_z^{(2)}\Big) I_z^{(k)}$ affects the pseudo-spin of the pair. In the pseudo-spin subspace we then obtain
\begin{equation}
    H_{int} = \sum_k (A_k^{(1)} - A_k^{(2)}) \hat{I_z} I_z^{(k)} + H_B,
\end{equation}
where $\hat{I}_z$ is the pseudo-spin operator and the sum is over all $k$ bath spins. The pair forms a decoherence-free subspace: only the difference in coupling of a bath spin $k$ causes noise on the pseudo-spin of the pair. For a single spin $A_k^{(2)} = 0$ and $\hat{I}_z$ becomes the single-spin operator $I_z$.

Following analyses by Anderson et al. \cite{AndersonP.W.;Weiss1953} and Klauder et al. \cite{KlauderJ.R.;Anderson1962}, we now take a classical approach to the noise in the pseudo-spin subspace. We model $\Delta Z(t)$ as an Ornstein-Uhlenbeck process with correlation function $\langle \Delta Z (t) \Delta Z (0) \rangle = b^2 \exp(-t/\tau_c)$ where $\tau_c$ is the correlation time of the bath due to the intra-bath dynamics $H_B$ and $b^2 = \frac{1}{4}\sum_k (A_k^{(1)} - A_k^{(2)})^2$ is the variance of the noise.

We calculate the distributions for $b$ numerically (see section \ref{b_distributions}). For nearest-neighbour pairs (like pairs A and B), we find a typical $b \sim 10$ Hz. For the parameters of pair C we find a larger value $b \sim 15$ Hz, consistent with the larger distance between the spins leading to less correlated noise and a less effective decoherence-free subspace. As a reference, for an individual $^{13}$C spin one has $b \sim 20$ Hz. For the correlation time when the NV spin is in $m_s = 0$, a typical value $\tau_c = 0.1$ s can be estimated from previous experiments \cite{Cramer2016}.

\subsection{Pair C and $m_s = -1$: no clock transition, a frozen core, quasi-static noise}
For pair C we have $Z \gg X$. For the NV in $m_s=-1$, the effect of the coupling $X$ becomes negligible, so that equation (S1) can be approximated as 
\begin{equation}
    H \approx (-Z + \Delta Z (t)) \hat{I}_z.
\end{equation}

In this case, $\Delta Z (t)$ directly and linearly modifies the eigenenergies. There is no coherence protection related to a clock transition. Additionally, in the $m_s = -1$ state, the NV magnetic field gradient creates a frozen core, in which nuclear spin flip-flops are suppressed \cite{Bradley2019,Zhong2015}. As a result, the dephasing time ($T_2^*$) is short compared to the correlation time of the noise $\tau_c$, and $\Delta Z (t)$ can be described as quasi-static. 

This regime leads to a Gaussian decay of $\exp(-b^2 \tau^2 /2)$ with $T_2^* = \frac{\sqrt{2}}{b}$ \cite{AndersonP.W.;Weiss1953}. Therefore, we can extract $b$ from the experimental data. The measured $T_2^*$ time of 18(1) ms yields $b_C = 2 \pi \cdot 12.5(7)$ Hz, consistent with the expected distribution for the inter-pair distance of pair C (Supplementary Fig. \ref{deltaZ}). The hypothesis that the noise can be treated as quasi-static is further corroborated by the fact that a large increase in coherence is observed for a spin echo ($T_2 = 0.3(2)$ s, Extended Data Fig. 3d).

In summary, for pair C and $m_s = -1$, we probe a regime where the clock transition is turned off, the decoherence-free subspace has a reduced influence (larger inter-pair distance) and the bath noise can be treated as quasi-static (frozen core and $T_2^* \ll \tau_c$). This regime and the resulting $T_2^*$ is similar as for an individual nuclear spin in the same environment \cite{Bradley2019}. No significant enhancement of coherence is obtained for the spin pair.    

\subsection{Pair A, B and $m_s = -1$: a detuned clock transition, a frozen core,  quasi-static noise}
For pair A, B we have $X \gg Z$. Additionally we typically have a situation in which $X \gg Z \gg \Delta Z (t)$. Taking only terms that contribute to dephasing, we can therefore approximate the Hamiltonian with the NV in $m_s = -1$ using a Taylor series expansion as 
\begin{equation}
    H = \omega_{-1}\hat{I}_x + \frac{Z}{\omega_{-1}} \Delta Z (t) \hat{I}_x,
\end{equation}
where $\omega_{-1} = \sqrt{X^2+Z^2}$. Similarly to the case of pair C, we expect a Gaussian decay shape, but now with  $T_2^* = \frac{\omega_{-1}}{Z}\frac{\sqrt{2}}{b}$. While the coupling $X$ now creates a clock transition, the system is detuned from the ideal point, because $Z \gg \Delta Z(t)$. As a result, the effect of the noise is reduced by a factor $\frac{\omega_{-1}}{Z} \approx 20$, consistent with the increase of $T_2^*$ of pair A and B compared to pair C.

The experimental data contains the decay of both pair A and B that are generally not equal. We extract two decay times from the Gaussian fit of the Fourier transform (Fig. 3a), obtaining $T_{2,A}^* = 0.26(2)$ s and $T_{2,B}^* = 0.39(6)$ s. For pair A with $Z_A = 2 \pi \cdot 130(1)$ this corresponds to $b_A = 2 \pi \cdot 13.9(2)$ Hz and for pair B with $Z_B = 2 \pi \cdot 91(2)$ Hz this corresponds to $b_B = 2 \pi \cdot 12.5(4)$ Hz. The values agree with typical values for $b$ (Supplementary Fig. \ref{deltaZ}).

In conclusion, in this case we probe a regime where the pair interaction $X$ creates a clock transition (as $X \gg Z$), but the system is detuned because $Z \gg \Delta Z(t)$. Additionally, the NV electron spin creates a frozen core and the noise can thus be treated as quasi-static. The resulting dephasing time is enhanced by a factor $\sqrt{X^2+Z^2}/Z$.

\subsection{Pair A, B and $m_s = 0$: clock transition, motional narrowing}

In $m_s = 0$, the noise cannot be treated as quasi-static anymore, because flip-flops between $^{13}$C spins occur more frequently. We therefore have to take into account the correlation time $\tau_c$ of the bath as well as its strength $b$. The Hamiltonian for a pair in the pseudo-spin subspace in $m_s = 0$ is
\begin{equation}
    H = X\hat{I}_x + \Delta Z(t)\hat{I}_z.
\end{equation}
For pair A, B it holds that $X \gg \Delta Z (t)$ for typical values of $\Delta Z (t)$ (Supplementary Fig. \ref{deltaZ}). We initially assume that the bath has no significant frequency components leading to direct transitions between the pair eigenstates ($X \gg 1/\tau_c$), but will come back to this effect at the end of the section. Then, following the analysis by Dobrovitski et al. \cite{Dobrovitski2009}, we can approximate the Hamiltonian as
\begin{equation}
    H = X \hat{I}_x + \frac{\Delta Z^2(t)}{2X}\hat{I}_x,
    \label{clock_state_equation}
\end{equation}
with $\Delta Z (t)$ a Gaussian distribution with variance $b^2 = \frac{1}{4}\sum_k (A_k^{(1)} - A_k^{(2)})^2$. The system now forms an effective clock transition and the noise term $\Delta Z (t)$ enters quadratically in the Hamiltonian.

This model can be solved analytically for the expectation value $\expval{S_z}$ of the pair pseudo-spin \cite{Dobrovitski2009}:
\begin{align}
\begin{split}
    \expval{S_z(t)} &= \frac{1}{2}\text{Re}[M(t)\exp(iXt)]\\
    [M(t)]^{-2} &= \exp(-Rt)[\cosh (Pt) + (R/P)\sinh (Pt)] - i \frac{b^2}{XP} \exp(-Rt)\sinh (Pt) \label{pair_solution_ms0}
\end{split}
\end{align}
where $P = \sqrt{R^2 - 2ib^2R/X}$ and $R = 1/\tau_c$ where $\tau_c$ is the correlation time of the bath.\\

Equation (\ref{pair_solution_ms0}) holds generally, but it is possible to consider three different regimes separately. These regimes are defined by the rate of the noise bath fluctuations $R$ compared to the effective coupling to the noise bath of $b^2/2X$. First, we consider a quasi-static bath. This leads to a non-exponential decay of the form \cite{Dobrovitski2003,Dobrovitski2009}
\begin{equation}
    M(t) = \Big(1+\Big(\frac{b^2t}{X}\Big)^2\Big)^{-\frac{1}{4}}.
    \label{quasi_static_bath}
\end{equation}

Second, the noise bath can be `slow' compared to the coupling strength to the noise: $R \ll b^2/X$. For short times, during which the bath is static, the decay follows equation (\ref{quasi_static_bath}). For longer times, during which slow bath dynamics have to be taken into account, the decay is of the form \cite{Dobrovitski2009}
\begin{equation}
    M(t) = \exp(-bt \sqrt{R/4X}).
    \label{slow_bath}
\end{equation}

Third, the bath dynamics can be fast compared to the magnitude of the noise: $R \gg b^2/X$. In this regime we can approximate the solution in equation (\ref{pair_solution_ms0}) as \cite{Dobrovitski2009}
\begin{equation}
    M(t) = \exp(i\frac{b^2t}{2X} - \frac{b^4t}{4X^2R}).
    \label{motional_narrowing_equation}
\end{equation}
The expected decay time of the Ramsey experiment is $T_2^* = \frac{4X^2R}{b^4}$. In this regime the pair coherence benefits from an effect that is similar to motional narrowing \cite{Dobrovitski2009}: $T_2^*$ linearly increases with the rate of fluctuations $R$. The result differs from the standard case of motional narrowing in that there is an additional frequency shift of $b^2/2X$ \cite{Dobrovitski2009}.\\

We now discuss which of these regimes governs the coherence of pair A and B. As shown in the previous section $b_A = 2 \pi \cdot 13.9(2)$ Hz and $b_B = 2 \pi \cdot 12.5(4)$ Hz. From the main text we know $X = 2 \pi \cdot 2080.9900(3)$ Hz (Fig. 3b). First we consider the expected dephasing times for slow baths. Equation (\ref{pair_solution_ms0}) is plotted in Supplementary Fig. \ref{equations_plotted_final} for pair A ($b_A = 2 \pi \cdot 13.9(2)$ Hz) with a correlation time of $\tau_c = 10$ s. From Supplementary Fig. \ref{equations_plotted_final} it is clear that a slow bath cannot explain the long inhomogeneous dephasing time observed in the main text ($T_2^* = 1.9(3)$ min, Fig. 3c).\\

\begin{figure}[h]
\includegraphics[scale=1.0]{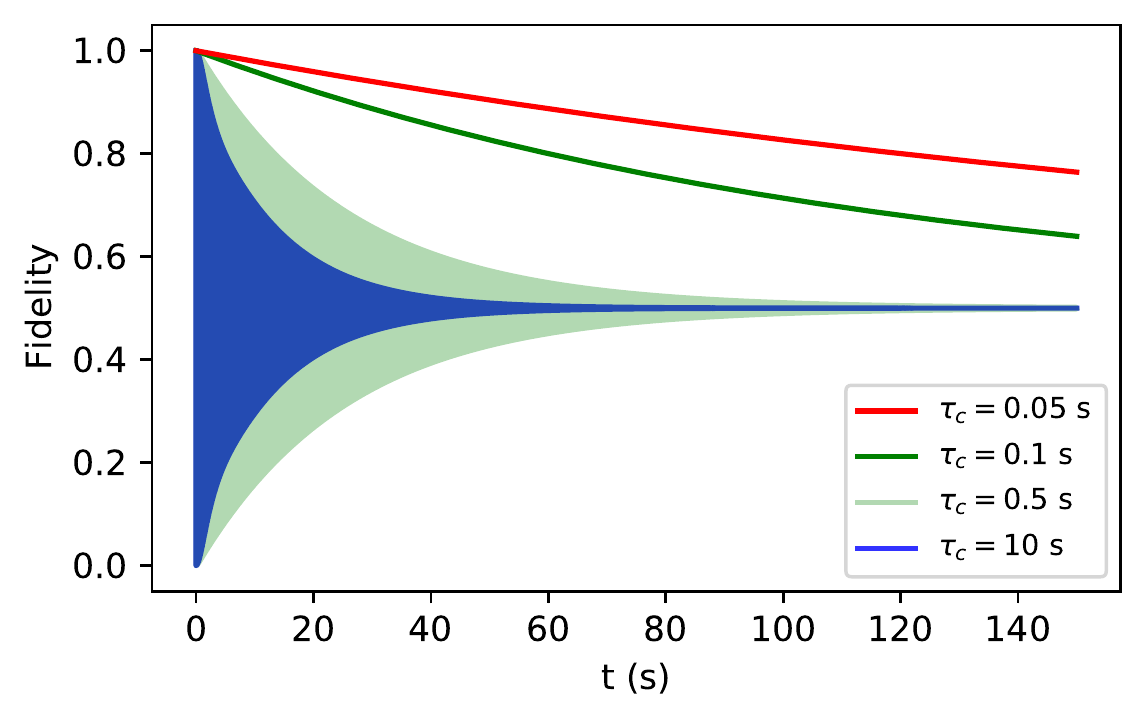}
\caption{\textbf{Dephasing-limited signals and envelopes for fast and slow baths.} The results for $\tau_c = 10$ s and $\tau_c = 0.5$ s are calculated from equation (\ref{pair_solution_ms0}). For $\tau_c = 0.1$ s and $\tau_c = 0.05$ s equation (\ref{motional_narrowing_equation}) holds and we therefore calculate the dephasing-limited envelope from equation (\ref{motional_narrowing_equation}). We have used $X = 2 \pi \cdot 2080.9900$ Hz and $b = 2 \pi \cdot 13.9$ Hz.}
\label{equations_plotted_final}
\end{figure}

Because typical measured relaxation times $T_1$ for individual $^{13}$C spins indicate $\tau_c = 0.1-0.3$ seconds \cite{Cramer2016}, one would indeed expect that the bath noise enters a fast regime: the fast bath condition $R \gg b^2/X$ is satisfied for bath correlation times of $\tau_c \lesssim 0.5$ s.

The envelopes for a fast bath are shown in Supplementary Fig. \ref{equations_plotted_final}. These results show that significantly longer dephasing times are expected in this regime compared to the slow bath regime. We conclude that the measured $T_2^* = 1.9(3)$ min can only be explained in this fast bath regime where an effect similar to motional narrowing further enhances the inhomogeneous dephasing time.

Since the coupling to the noise of pair A ($b_A$) and B ($b_B$) were determined above, the bath fluctuation rate can be estimated under the assumption that dephasing is limiting. Using $T_2^* = 1.9(3)$ min and $X = 2 \pi \cdot 2080.9900(3)$ Hz we obtain $R_A = 10(2)$ Hz and $R_B = 6(2)$ Hz. These values are consistent with previously measured values for $T_1$ of individual $^{13}$C spins with the NV electron spin in $m_s = 0$ \cite{Cramer2016}.

Finally, we consider relaxation as a potential mechanism limiting the dephasing time. Spectral components around frequency $X$ could lead to direct transitions between the antiparallel pair eigenstates ($[\ket{\Uparrow}+\ket{\Downarrow}]/\sqrt{2} \leftrightarrow [\ket{\Uparrow}-\ket{\Downarrow}]/\sqrt{2} $). In the regime of a fast bath this can be treated analytically and introduces a factor that multiplies equation (\ref{pair_solution_ms0}) by $\exp(-b^2Rt/2X^2)$ \cite{Dobrovitski2009}. If this type of relaxation is dominant we would obtain $T_2^* = \frac{2X^2}{b^2R}$. The identified values for $b_A$, $b_B$, $X$ and $T_2^*$ give $R_A = 400(75)$ Hz and $R_B = 500(100)$ Hz, which is inconsistent with previously measured single $^{13}$C spin $T_1$ values \cite{Cramer2016} and with the typical $^{13}$C-$^{13}$C couplings for this $^{13}$C concentration. We conclude that such relaxation within the pseudo-spin subspace is unlikely to contribute to the observed coherence curves.

In summary, the long observed dephasing times for pair A, B are the result of three different physical effects working together. First, the effective noise $b$ is reduced because correlated noise does not affect the spin pair, i.e. the pair forms a decoherence-free subspace (DFS). Second, since $X \gg \Delta Z (t)$, the pair pseudo-spin forms a clock transition (equation (\ref{clock_state_equation})). It is therefore only second-order sensitive to noise following $\Delta Z^2 (t) / 2X$. Third, the DFS and clock transition alone are not sufficient to explain a dephasing time of $1.9(3)$ min. Only in the regime of a fast bath, in which the pair benefits from an effect similar to motional narrowing, can such dephasing times be realized.

\subsection{Pair A, B and $m_s=0$: other decoherence mechanisms}

Above, we show how a combination of three effects strongly suppresses dephasing for the spin pairs. This strong suppression of dephasing is a necessary condition to obtain the long $T_2^*$ times observed. However, it does not imply that the obtained $T_2^*$ times and decay curves are explained by and purely limited by dephasing. Indeed, the decay envelope observed for pairs A and B (Fig. 3c) deviates from the simple exponential decay obtained in equation (\ref{motional_narrowing_equation}). In particular, the fit yields a decay curve following $e^{-(t/T_2^*)^n}$ with $n=0.23(2)$ (Methods) and the data suggests additional features in the decay shape that are not captured by the fit curve. These observations indicate that other mechanisms, like coherent interactions with the bath or relaxation of the pair spins due to flip-flops with the bath spins are contributing to decoherence. Such effects strongly depend on the microscopic environment of the pairs and are challenging to treat generally. Future research could aim to understand the microscopic environment of the pairs and determine the mechanisms limiting the observed coherence times.

\subsection{Pair C and $m_s = 0$}
The analysis for pair C is analogous to the above analysis. However, there is an important difference between pair A, B and pair C. Namely, the dipolar coupling $X$ is an order of magnitude smaller for pair C. Since the effective noise strength for pair C ($b_C = 2 \pi \cdot 12.5(7)$ Hz) is similar to the noise for pair A ($b_A = 2 \pi \cdot 13.9(2)$ Hz) and pair B ($b_B = 2 \pi \cdot 12.5(4)$), the approximation of a fast bath cannot be easily made. Therefore we are in an intermediate regime where the full expression in equation (\ref{pair_solution_ms0}) describes the dephasing. The pair C coherence in $m_s = 0$ therefore still benefits from the decoherence-free subspace and clock transition, but to a lesser degree from motional narrowing.

\subsection{Pair C: spin echo $T_2$, relaxation time $T_1$}

For pair C, spin echo ($T_2$) and relaxation ($T_1$) measurements in both electron states were taken (Extended Data Fig. 3). When the NV electron spin is in $m_s = -1$, the NV spin creates a hyperfine field gradient that slows down spin flip-flops in the bath (a frozen core \cite{Bradley2019,Zhong2015}). The noise a $^{13}$C pair experiences is therefore expected to be quasi-static. Given quasi-static noise, a spin echo is expected to increase coherence. This is in agreement with the marked increase in coherence time: $T_2 = 0.3(2)$ s (Extended Data Fig. 3d). Similarly, the frozen core also suppresses flip-flops involving one of the $^{13}$C spins of the pair, leading to long relaxation times. For pair C we do indeed find a relaxation time $T_1 \gg 1$ s, comparable to that of single $^{13}$C spins \cite{Bradley2019}.

When the NV electron spin is in $m_s = 0$, $^{13}$C spins are no longer detuned and flip-flops can become limiting for spin coherence. For pair C we find a spin echo time of $T_2 = 0.7(1)$ s. The relaxation time is $T_1 = 0.9(2)$ s or $T_1 = 3.6(7)$ s depending on the initial state. This suggests that the relaxation mechanism is dependent on initialisation in the singlet ($(\ket{\Uparrow} - \ket{\Downarrow})/\sqrt{2}$) or triplet state ($(\ket{\Uparrow} + \ket{\Downarrow})/\sqrt{2}$).
Furthermore these results indicate that the coherence of pair C may be limited by relaxation.

\newpage

\section{Decoherence-free subspace: distributions for $b$} \label{b_distributions}
\begin{figure}[h]
\includegraphics[scale=0.4]{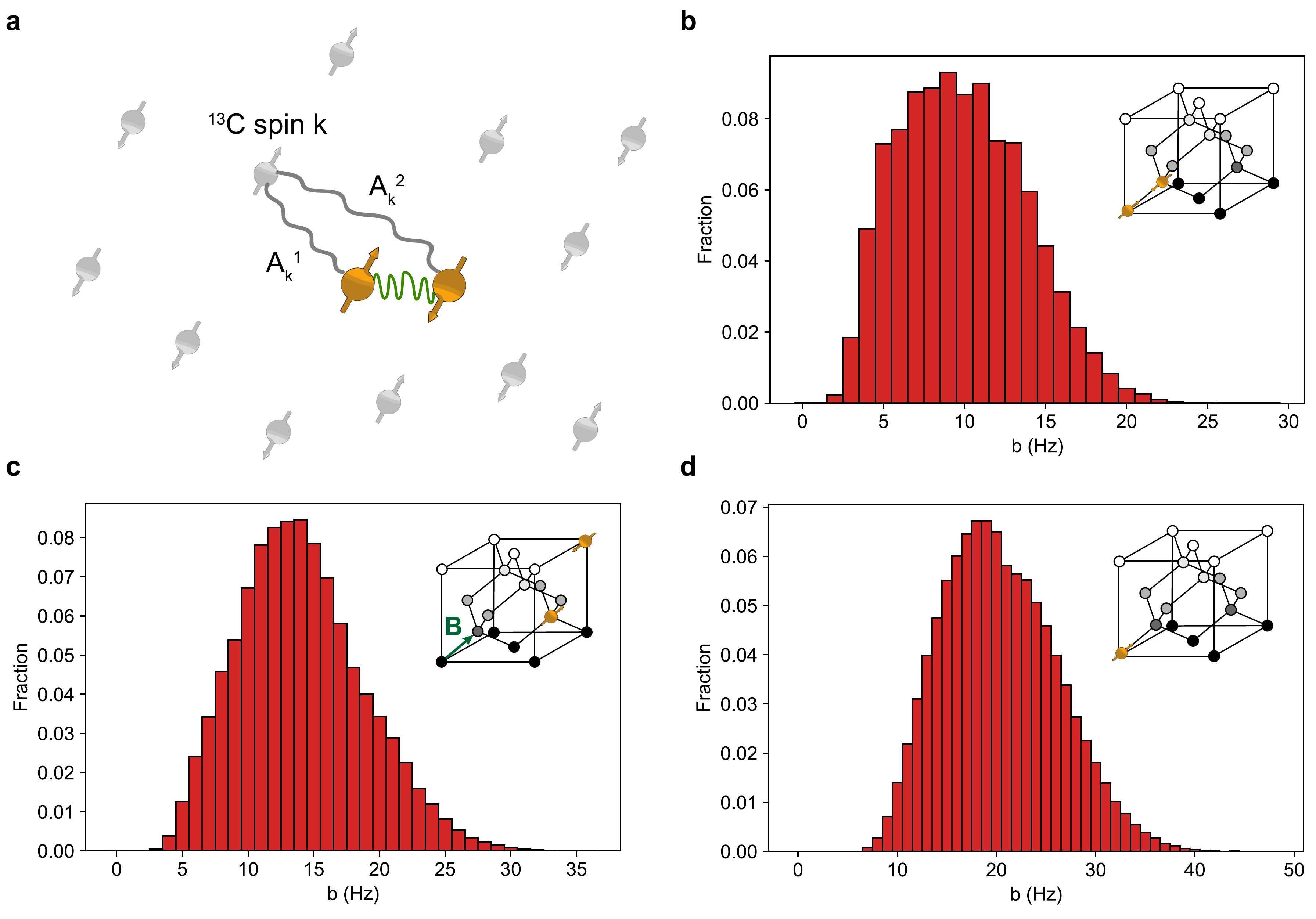}
\caption{\textbf{Distributions of the noise strength $b$ for typical baths.} \textbf{a.} Schematic of the situation. A pair is surrounded by $k$ bath spins that each create a magnetic field difference $A_k^{(1)} - A_k^{(2)}$ on the pair. Assuming Gaussian noise, we obtain $b$ from $b^2 = \frac{1}{4}\sum_k (A_k^{(1)} - A_k^{(2)})^2$. \textbf{bcd.} Distribution of $b$ for $10^5$ generated baths for the parameters of pair A and B (b), for the parameters of pair C (c) and for a single spin (d). Strongly coupled spins ($\abs{A_k^{(1)}} > 50$ Hz or $\abs{A_k^{(2)}} > 50$ Hz) have been excluded.}
\label{deltaZ}
\end{figure}

The noise $\Delta Z (t)$ on a spin pair originates from the surrounding $^{13}$C spins. As a pair is only sensitive to field gradients (a decoherence-free subspace), distant external noise sources can generally be neglected. There are $k$ bath spins that each create a field difference $A_k^{(1)} - A_k^{(2)}$ on the pair (Supplementary Fig. \ref{deltaZ}a). As outlined above, we model $\Delta Z(t)$ as an Ornstein-Uhlenbeck process with a variance $b^2 = \frac{1}{4}\sum_k (A_k^{(1)} - A_k^{(2)})^2$. The question that we address in this section is what $b$ is for typical spin baths.

We numerically generate $10^5$ different baths ($1.1 \%$ $^{13}$C abundance) surrounding a pair (Supplementary Fig. \ref{deltaZ}b,c) or single spin (Supplementary Fig. \ref{deltaZ}d) in a volume of $15 \times 15 \times 15$ unit cells. For each bath we calculate $b^2 = \frac{1}{4}\sum_k (A_k^{(1)} - A_k^{(2)})^2$ but exclude spins with $\abs{A_k^{(1)}} > 50$ Hz or $\abs{A_k^{(2)}} > 50$ Hz, i.e. we exclude strongly coupled spins for which the system would not be a well-defined spin pair anymore. The expectation is that the closer the spins of the pair are, the more correlated the noise and the smaller $b$ is.

The result for a nearest neighbour pair oriented along the magnetic field (like pair A or B) is shown in Supplementary Fig. \ref{deltaZ}b. We find a mean of $10$ Hz and a standard deviation of $4$ Hz. For the parameters of pair C (Supplementary Fig. \ref{deltaZ}c) we find a mean of $14$ Hz and a standard deviation of $5$ Hz. Lastly for an individual $^{13}$C spin (Supplementary Fig. \ref{deltaZ}d) we find a mean of $20$ Hz and a standard deviation of $6$ Hz. A decrease in the effective noise is observed for the pairs compared to an individual spin. Furthermore, the closer the pair spins are, the smaller the effective noise is.

\newpage

\section{Decoherence-free subspace and external noise} \label{effect_external_noise}
The decoherence-free subspace makes pairs nearly immune to noise from distant sources. In this section we consider the effect of two such external sources. For single $^{13}$C spins in the same sample typical inhomogeneous dephasing times of 10 ms are observed \cite{Bradley2019}, which sets a bound on the noise strength. If we take the extreme case that all noise comes from an external source, i.e. not from the spin bath, this gives an upper limit of the noise magnitude $b = \sqrt{2}/T_2^* = 2 \pi \cdot 22.5$ Hz. Since we are interested in an order of magnitude estimate, we take $b \sim 2 \pi \cdot 10^2$ Hz, corresponding to magnetic field fluctuations of $\delta B \sim 10^{-5}$ T. Now we consider that these fluctuations originate from either the on-chip MW line that we use to apply microwaves or from the external magnets that we use to apply a magnetic field.

\subsection{Microwave line}

We approximate the microwave line as an infinite wire that generates a field at a distance $r$ from the wire of magnitude $B(r) = \mu_0 I /2 \pi r$ where $\mu_0$ is the vacuum permeability and $I$ the current through the wire. Given $r \sim 10$ \textmugreek m and $\delta B \sim 10^{-5}$ T, we obtain $I = 2 \pi r \delta B / \mu_0 \sim 10^{-4}$ A. 
Now we turn to the effect of this noise on a decoherence-free subspace formed by a $^{13}$C pair. The positions of the pair spins are $r_a = 10$ \textmugreek m and $r_b = 10$ \textmugreek m $+$ $a_0$ where for $a_0$ we take a conservative value of $\sim 10^{-9}$ m. Given $I = 10^{-4}$ A, the MW line would add a field difference to the decoherence-free subspace of $\Delta B = B(r_a) - B(r_b) = \frac{\mu_0 I}{2 \pi}(\frac{1}{r_a} - \frac{1}{r_b}) \sim 10^{-10}$ T. That corresponds to $\sim 10^{-3}$ Hz which has a negligible effect on the coherence.

\subsection{Magnet}
The external magnetic field comes from a cylindrical permanent magnet. We calculate the effect of that field ($\sim 0.04$ T) on the decoherence-free subspace of a pair. From the above we know that the maximum field fluctuations are on the order of $\delta B \sim 10^{-5}$ T. We consider a magnet with an NV-magnet distance $r \sim 10^{-2}$ m, the radius of the magnet $R = 5$ mm, the length $L = 5$ mm and the remanence field $B_r = 1.5$ T. To calculate the magnetic field at $r$ we use
\begin{equation}
    B(r) = \frac{B_r}{2}\Big(\frac{L+r}{\sqrt{R^2+(L+r)^2}}-\frac{r}{\sqrt{R^2+r^2}}\Big)
\end{equation}
At $r \sim 10^{-2}$ m the magnetic field is $\sim 0.04$ T.
The expected effect of the permanent field on the decoherence-free subspace is then $B(r_a) - B(r_b) \sim 10^{-8}$ T or $10^{-1}$ Hz. We have used $r_a = 1$ cm and $r_b = 1$ cm $+$ $a_0$ where for $a_0$ we take a conservative value of $10^{-9}$ m. This is a constant field difference (c.f. $Z$) added to the pair. However, the field fluctuations are $\delta B \sim 10^{-5}$ T, more than three orders of magnitude less than the permanent field of $\sim 0.04$ T. The influence of $\delta B$ on the decoherence-free subspace is therefore $< 1$ mHz which has a negligible effect on the coherence.

\newpage

\section{Expected number of nearest-neighbour pairs per NV}
In this section we address how many nearest neighbour pairs with similar $Z$ as pair A and B one would expect surrounding a typical NV center. To that end we generate $10^4$ different $^{13}$C baths with $1.1\%$ abundance surrounding an NV center in a volume of $15 \times 15 \times 15$ diamond unit cells. For every generated bath, we look for the nearest neighbour pairs along the magnetic field axis and calculate the hyperfine field difference $Z$ due to the NV, assuming a dipolar NV-$^{13}$C interaction. Then we estimate a controllable region of $2 \pi \cdot 50<Z< 2 \pi \cdot 500$ Hz. The upper bound comes from the required condition $X \gg Z$ and the (approximate) lower bound is a limit due to the detrimental effect of electron dephasing for a large number of dynamical decoupling pulses. Additionally for smaller $Z$ resolving a pair from the background bath of pairs with small $Z$ is expected to be challenging. For every generated bath, we determine how many pairs satisfy this condition and plot the result in Supplementary Fig. \ref{number_of_pairs}. The expected number of such nearest neighbour pairs per NV is $1\pm 1$. Moreover, more than $70 \%$ of simulated NVs host at least one nearest-neighbour pair, indicating that such pairs can commonly be found.

In the above we only consider nearest-neighbour pairs along the magnetic field axis. Other pairs with smaller $X$ and larger $Z$ can also be detected and controlled (pair C, see Methods). In Supplementary Table \ref{X_table} we show the ten largest values of $X$ with their corresponding occurrence and vector $\mathbf{r}$ between the two $^{13}$C spins.

\begin{figure}[h]
\includegraphics[scale=0.7]{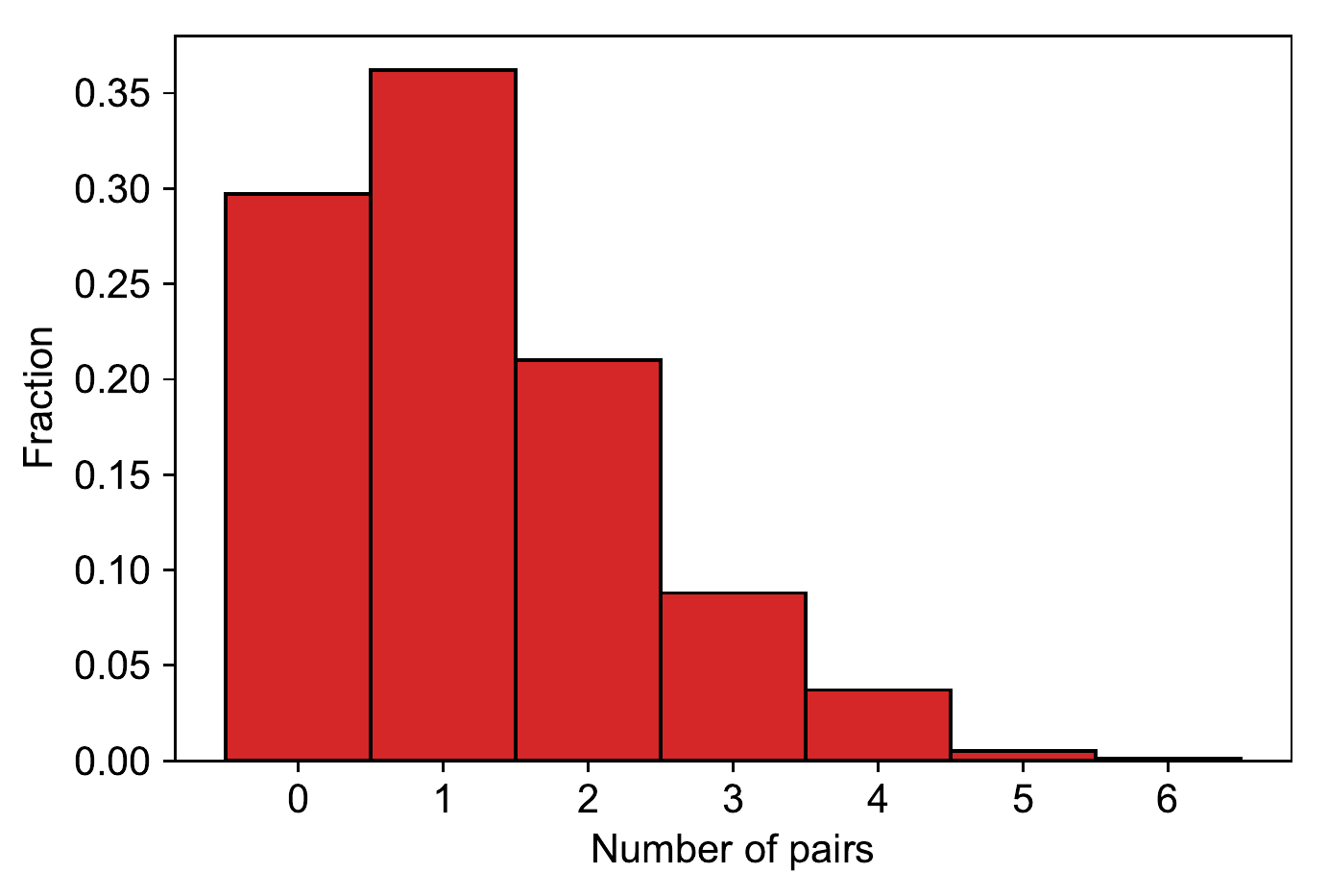}
\caption{\textbf{Expected number of nearest neighbour pairs per NV}. The expected number of pairs with $2 \pi \cdot 50<Z< 2 \pi \cdot 500$ Hz for a volume of $15 \times 15 \times 15$ unit cells surrounding an NV center. We estimate that pairs with a $Z$ within the indicated region to be controllable with high fidelity. When $Z$ is larger, it becomes comparable to $X$ and the control methods presented in this paper do not hold anymore. When $Z$ is much smaller, more pulses are required in the dynamical decoupling sequence resulting in more electron dephasing, and resolving the pair from the background bath of pairs becomes more challenging.}
\label{number_of_pairs}
\end{figure}

\begin{table}[h]
\centering
\begin{tabular}{|l|l|l|}
\hline
$X$ (Hz) & occurrence & $\mathbf{r}$ (in units $a_0/4$)\\ \hline
2062 & 1 &[1,1,1]\\ \hline
687 & 3 &[-1,1,-1]\\ \hline
237 & 12 &[$\pm$2,$\pm$2,0] and [$\mp$2,$\pm$2,0]  \\ \hline
187 & 3 &[-1,-1,-3] \\ \hline
134 & 3 &[1,1,-3] \\ \hline
102 & 3 &[1,3,3] \\ \hline
76.38 & 1 &[-3,-3,-3] \\ \hline
75.95 & 6 &[$\pm$4,$\pm$2,$\pm$2] \\ \hline
61 & 6 &[-3,-1,3] \\ \hline 
46 & 6 &[$\mp$4,$\pm$2,$\pm$2] \\ \hline
\end{tabular}
\caption{The occurrence of a given coupling $X$ when fixing one of the $^{13}$C spins of the pair and moving the other around the lattice. The vector between the two $^{13}$C spins of the pair is given for each $X$ coupling in units of $a_0/4$ where $a_0$ is the lattice constant of diamond. All permutations of the entries of $\mathbf{r}$ give the same coupling $X$.} 
\label{X_table}
\end{table}

\newpage

\section{Pair initialisation and readout calibration}
In this section we describe the optimization of the parameters used for initialisation and single-shot measurements. The most important trade-off lies in the number of repetitions of the measurement sequences. On the one hand, increasing the number of repetitions improves the fidelity because the different states can be distinguished better and the effect of the NV electron spin dephasing is diminished. On the other hand, the pair spin state decoheres during the measurement, limiting the maximum number of repetitions. Therefore, there is an optimum in the number of repetitions and the corresponding decision thresholds used. 

We first describe our approach to optimize the parameters in general. We will call the two states that we want to optimally distinguish $\ket{a}$ and $\ket{b}$. In an initialisation step, we generally use $k$ repetitions and we record the number of counts ($m_s=0$ outcomes) as $N(k)$. This initialisation step is then defined by $N(k) > N_a$ and $N(k) < N_b$ where $N_a$ and $N_b$ are the thresholds set (red lines in Fig. 2f,g). In case there is a two-step initialisation process (see e.g. Fig. 2e) we denote the number of counts in the first step as $N(p)$ with condition $N(p) > N_0$ where $N_0$ is the threshold set.
% Even if the initialisation is a two-step process (Fig. 2ef), the last step is used to either initialise in $\ket{a}$ or $\ket{b}$.
In the readout step, we use $m$ repetitions and obtain $N(m)$ counts. Two histograms are obtained (see e.g. Fig. 2f), one corresponding to each initialised state. To optimally distinguish these states, we sweep a threshold $T$ (see e.g. Fig. 2g) and obtain the combined initialisation and readout fidelity as
\begin{equation}
    F = \frac{F_{\ket{a}} + F_{\ket{b}}}{2} = \frac{1}{2}P(N(m) \geq T|N(k) > N_a \wedge N(p) > N_0) + \frac{1}{2}P(N(m) < T|N(k) < N_b \wedge N(p) > N_0).
    \label{fidelity_calculation}
\end{equation}

We then optimize the fidelity for the number of repetitions $m$, the measurement decision threshold $T$, and the initialisation thresholds $N_a$, $N_b$ and if applicable $N_0$. For the initialisation, the number of repetitions $k$ is not as important. The main trade-off now lies in the initialisation thresholds: we pick values that balance the resulting fidelity and the success probability (experimental rate). Namely, the stricter the threshold is, the higher the fidelity but the lower the experimental rate.

\newpage

\section{Calibration of the spin measurement for pair A and B}
For pair A and B we calibrate the spin measurement to optimally distinguish $\ket{a} = \ket{\Uparrow}\ket{\Uparrow}$ and $\ket{b} = \ket{\Downarrow}\ket{\Downarrow}$. The sequence is shown in Supplementary Fig. \ref{pair4_spin_calib}a. First, we initialise the parity of pair A and B with $p = 20$ repetitions and a threshold of $N_0 = 12$. To initialise the two states of interest we then use $k = 30$ spin readouts and set thresholds of $N_a = 25$ and $N_b = 3$ counts.\\
For a varying number of readouts $m$ we now determine the optimal threshold and corresponding fidelity as in Fig. 2g in the main text. In Supplementary Fig. \ref{pair4_spin_calib}b, we plot the average fidelity and corresponding optimal threshold for a varying number of readouts. The optimum is at $m = 35$ readouts with a threshold of $T = 16$. We obtain a combined initialisation and readout fidelity of $F = 98.4(7) \%$. For the results in the main text we use $m = 30$ and $T = 14$ which gives, within error, the same fidelity ($F = 98.2(7) \%$).\\

\begin{figure}[h]
\includegraphics[scale=0.4]{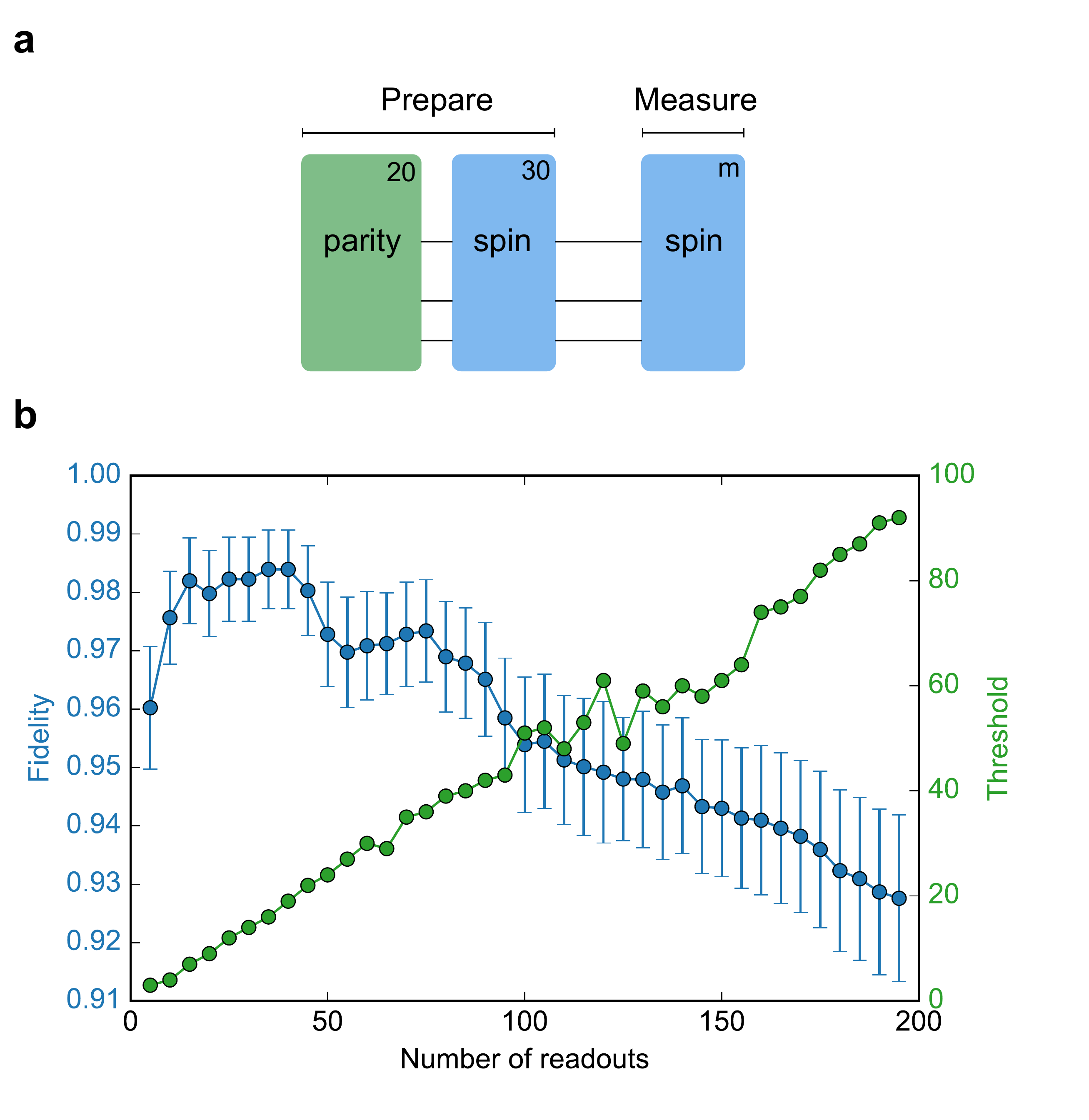}
\caption{\textbf{Spin measurement calibration for pair A and B.} \textbf{a.} The sequence used to calibrate the spin readout. First we select on $>14/20$ counts in 20 parity readouts to initialise the pairs in $\frac{1}{2}\ket{\Uparrow}\ket{\Uparrow}\bra{\Uparrow}\bra{\Uparrow} + \frac{1}{2}\ket{\Downarrow}\ket{\Downarrow}\bra{\Downarrow}\bra{\Downarrow}$. Then we either initialise the pairs in $\ket{\Uparrow}\ket{\Uparrow}$ ($>25/30$) or $\ket{\Downarrow}\ket{\Downarrow}$ ($<3/30$) with 30 spin readouts. Finally the spin is read out using $m$ spin readouts. \textbf{b.} We vary the number of readouts $m$ and find the optimal threshold for each $m$. The average fidelity (equation (\ref{fidelity_calculation})) is plotted against the number of readouts $m$ and the corresponding optimal threshold is indicated for that number of readouts. The initial increase of fidelity with the number of repetitions originates from the limiting effect of dephasing of the NV electron spin during the interaction sequence. For a small number of repetitions, this dephasing limits the measurement fidelity. But with more repetitions the conditional histograms (Fig. 2f) separate and NV electron dephasing plays a limited role. For a large number of repetitions a slow decay in the fidelity is observed, which is due to decoherence of the pair spin during the measurement sequence.}
\label{pair4_spin_calib}
\end{figure}

\newpage

\section{Calibration of the parity measurement for pair A and B}
\begin{figure}[h]
\includegraphics[scale=0.4]{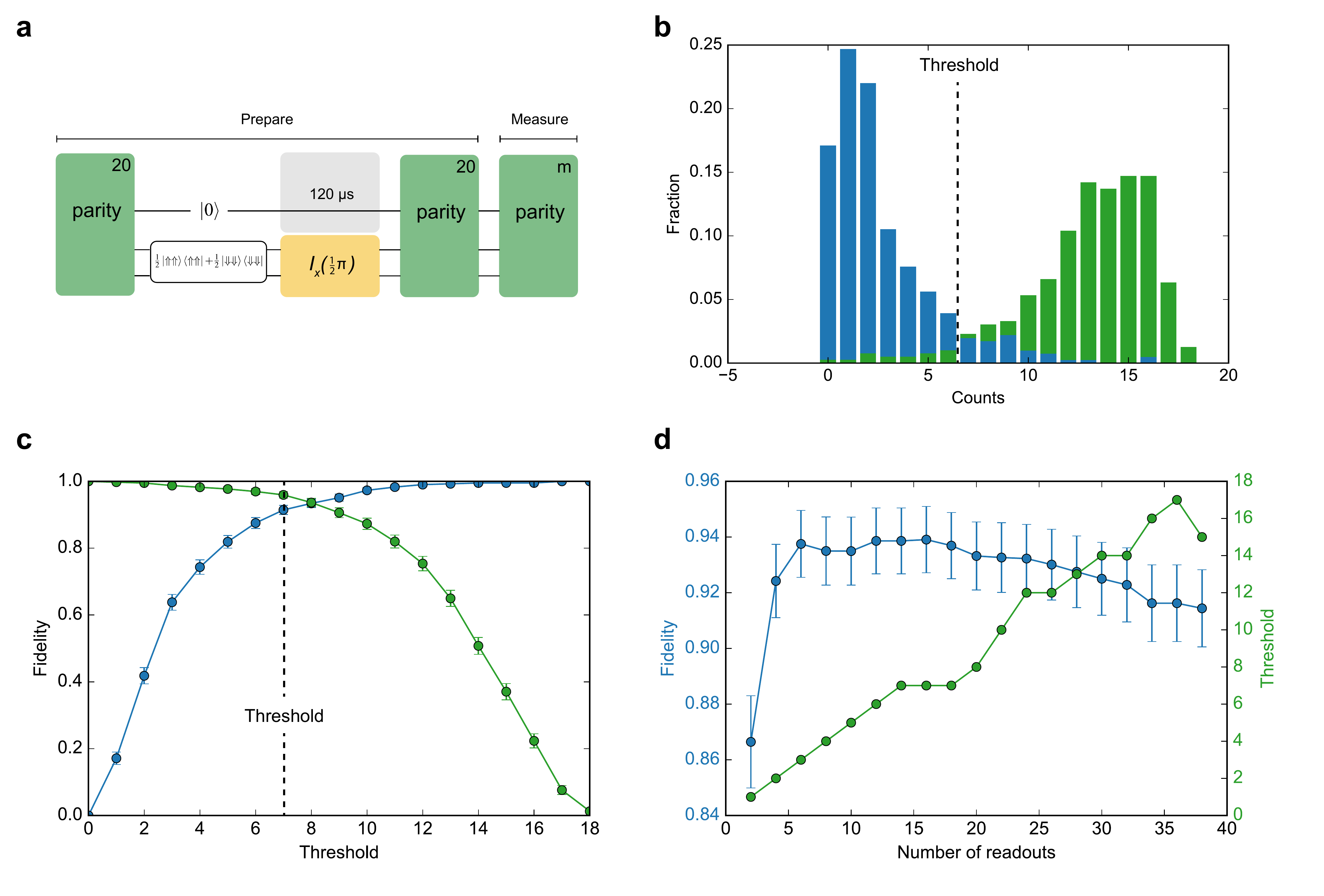}
\caption{\textbf{Parity measurement calibration of pair A and B} \textbf{a.} The sequence used to calibrate the pair A and B parity measurement. We select on $>14/20$ counts in the first 20 readouts and apply a $\pi/2$ around $x$ to obtain a mixed state in the antiparallel subspace. Then another 20 parity readouts are used to initialise pair A and B in even parity ($>15/20$) or odd parity ($<2/20$) states. Finally $m$ parity readouts are performed that we aim to calibrate. \textbf{b.} Conditional histograms for $\ket{a}$ (green) and $\ket{b}$ (blue) for $m = 18$ readouts. The optimal threshold of $T = 7$ is indicated. \textbf{c.} Combined initialisaton and readout fidelity of the individual states $\ket{a}$ (green) and $\ket{b}$ (blue) for $m = 18$ readouts. \textbf{d.} We vary the number of readouts $m$ and find the optimal threshold for each $m$. The average fidelity (equation (\ref{fidelity_calculation})) is plotted against the number of readouts $m$ and the corresponding optimal threshold is indicated for that number of readouts.}
\label{pair4_parity_calib}
\end{figure}

For pair A and B the parity measurement is optimized to distinguish the pair parity. To calibrate the measurement, we use the states $\ket{a} = \frac{1}{2}\ket{\Uparrow}\ket{\Uparrow}\bra{\Uparrow}\bra{\Uparrow} + \frac{1}{2}\ket{\Downarrow}\ket{\Downarrow}\bra{\Downarrow}\bra{\Downarrow}$ and $\ket{b} = \frac{1}{2}\ket{\Uparrow}\ket{\Downarrow}\bra{\Uparrow}\bra{\Downarrow} + \frac{1}{2}\ket{\Downarrow}\ket{\Uparrow}\bra{\Downarrow}\bra{\Uparrow}$. To prepare these two states, we use a two-step initialisation (Supplementary Fig. \ref{pair4_parity_calib}a). First, we use $p = 20$ parity initialisations with a threshold $N_0 = 14$ to obtain $\ket{a}$. Then we apply a $\pi/2$ pulse around $x$ to obtain a fully mixed state in the antiparallel subspace ($\{\ket{\Uparrow \Uparrow},\ket{\Uparrow \Downarrow},\ket{\Downarrow \Uparrow},\ket{\Downarrow \Downarrow}\}$). This first step of the initialisation has reduced the Bloch sphere in Fig. 2d to just contain antiparallel states (the pseudo-spin states), improving subsequent initialisation steps. The second step of the initialisation uses $k = 20$ readouts with initialisation thresholds of $N_a = 15$ and $N_b = 2$. We set the number of decoupling elements for the parity sequence to $N_p = 20$ in Fig. 2b. This is less than twice the decoupling elements for the spin sequence ($N_s = 14$ in Fig. 2a). The optimal interaction time is smaller than a $\pi$-rotation due to the loss of NV electron spin coherence with increasing $N_p$.

In Supplementary Fig. \ref{pair4_parity_calib}b the histograms for $\ket{a}$ (green) and $\ket{b}$ (blue) are shown for $m = 18$ readouts. We use equation (\ref{fidelity_calculation}) to obtain the fidelity of the individual states (Supplementary Fig. \ref{pair4_parity_calib}c) and the average fidelity while sweeping the threshold $T$. This process is repeated for a varying number of readouts $m$. In Supplementary Fig. \ref{pair4_parity_calib}d the average fidelity is shown as a function of the number of readouts $m$. The optimum is found for $m = 16$ readouts with a threshold $T = 7$. We obtain a combined initialisation and readout fidelity of $F = 94(1) \%$. For the results in the main text we use $m = 18$ and $T = 7$ which gives, within error, the same fidelity.\\

\newpage

\section{Calibration of the spin measurement for pair C}
\begin{figure}[h]
\includegraphics[scale=0.4]{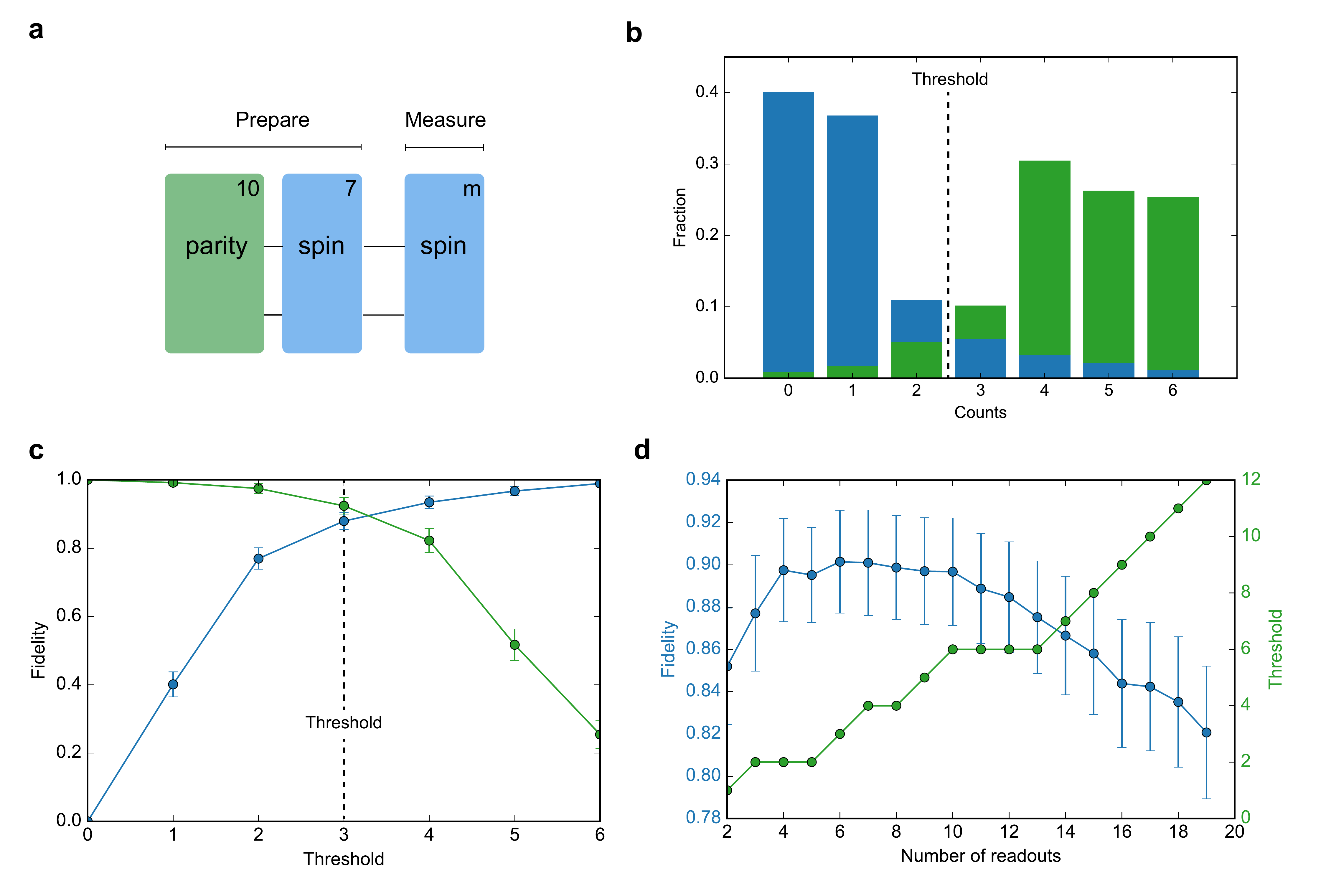}
\caption{\textbf{Spin readout calibration of pair C.} \textbf{a.} The sequence used to calibrate the pair C spin readout. We select on $>9/10$ in the 10 parity readouts to obtain a mixed state in the antiparallel subspace. Then we initialise $\frac{1}{\sqrt{2}}(\ket{\Uparrow} + \ket{\Downarrow})$ ($>6/7$) or $\frac{1}{\sqrt{2}}(\ket{\Uparrow} - \ket{\Downarrow})$ ($<1/7$). Finally we do $m$ spin readouts that we calibrate to optimally distinguish these states. \textbf{b.} Conditional histograms for $\frac{1}{\sqrt{2}}(\ket{\Uparrow} + \ket{\Downarrow})$ (green) and $\frac{1}{\sqrt{2}}(\ket{\Uparrow} - \ket{\Downarrow})$ (blue). The optimal threshold of $T = 3$ is indicated. \textbf{c.} Combined initialisation and readout fidelity of $\frac{1}{\sqrt{2}}(\ket{\Uparrow} + \ket{\Downarrow})$ (green) and $\frac{1}{\sqrt{2}}(\ket{\Uparrow} - \ket{\Downarrow})$ (blue) for $m = 6$ readouts. \textbf{d.} The average fidelity (equation (\ref{fidelity_calculation})) is plotted as a function of the number of readouts $m$ and the corresponding optimal threshold for the given number of readouts.}
\label{pair5_spin_calib}
\end{figure}

For pair C we optimize the spin readout to optimally distinguish $\ket{a} = \frac{1}{\sqrt{2}}(\ket{\Uparrow} + \ket{\Downarrow})$ and $\ket{b} = \frac{1}{\sqrt{2}}(\ket{\Uparrow} - \ket{\Downarrow})$ (Extended Data Fig. 2). To initialise these states, we first initialise the pair in the antiparallel subspace and then initialise the spin state (Supplementary Fig. \ref{pair5_spin_calib}a). We set $N_0 = 9$ in the $p = 10$ parity readouts and set $N_a = 6$ and $N_b = 1$ for the $k = 7$ spin readouts.

In Supplementary Fig. \ref{pair5_spin_calib}b the histograms for $\ket{a}$ (green) and $\ket{b}$ (blue) are shown for $m = 6$ readouts. We calculate the combined initialised and readout fidelity of $\ket{a}$ and $\ket{b}$ using equation (\ref{fidelity_calculation}) for varying thresholds $T$ (Supplementary Fig. \ref{pair5_spin_calib}c). In Supplementary Fig. \ref{pair5_spin_calib}d we vary the number of readouts $m$ and plot it against the average fidelity. The indicated threshold is the one that gives the maximum average fidelity. The optimum is found for $m = 6$ with a threshold $T = 3$. We obtain a combined initialisation and readout fidelity of $90(2) \%$.\\

\newpage

\section{Calibration of the parity measurement for pair C}
For pair C, the parity measurement distinguishes between the parallel and the antiparallel subspace (Extended Data Fig. 2). We calibrate the parity measurement to optimally distinguish $\ket{a} = \frac{1}{2}\ket{\Uparrow}\bra{\Uparrow} + \frac{1}{2}\ket{\Downarrow} \bra{\Downarrow}$ and $\ket{b} = \frac{1}{2}\ket{\uparrow \uparrow}\bra{\uparrow \uparrow} + \frac{1}{2}\ket{\downarrow \downarrow}\bra{\downarrow \downarrow}$. To initialise the subspace we use $k = 10$ parity readouts and set $N_a = 9$ and $N_b = 1$ (Supplementary Fig. \ref{pair5_parity_calib}a).

In Supplementary Fig. \ref{pair5_parity_calib}b the histograms for $\ket{a}$ (green) and $\ket{b}$ (blue) are shown for $m = 16$ readouts. We calculate the combined initialisation and readout fidelity of $\ket{a}$ and $\ket{b}$ using equation (\ref{fidelity_calculation}) for varying thresholds $T$ (Supplementary Fig. \ref{pair5_parity_calib}c). In Supplementary Fig. \ref{pair5_parity_calib}d we vary the number of readouts $m$ and plot it against the average fidelity. The indicated threshold is the one that gives the maximum average fidelity. The optimum is found for $m = 16$ with a threshold $T = 7$. We obtain a combined initialisation and readout fidelity of $95.9(6) \%$.\\

\begin{figure}[h]
\includegraphics[scale=0.4]{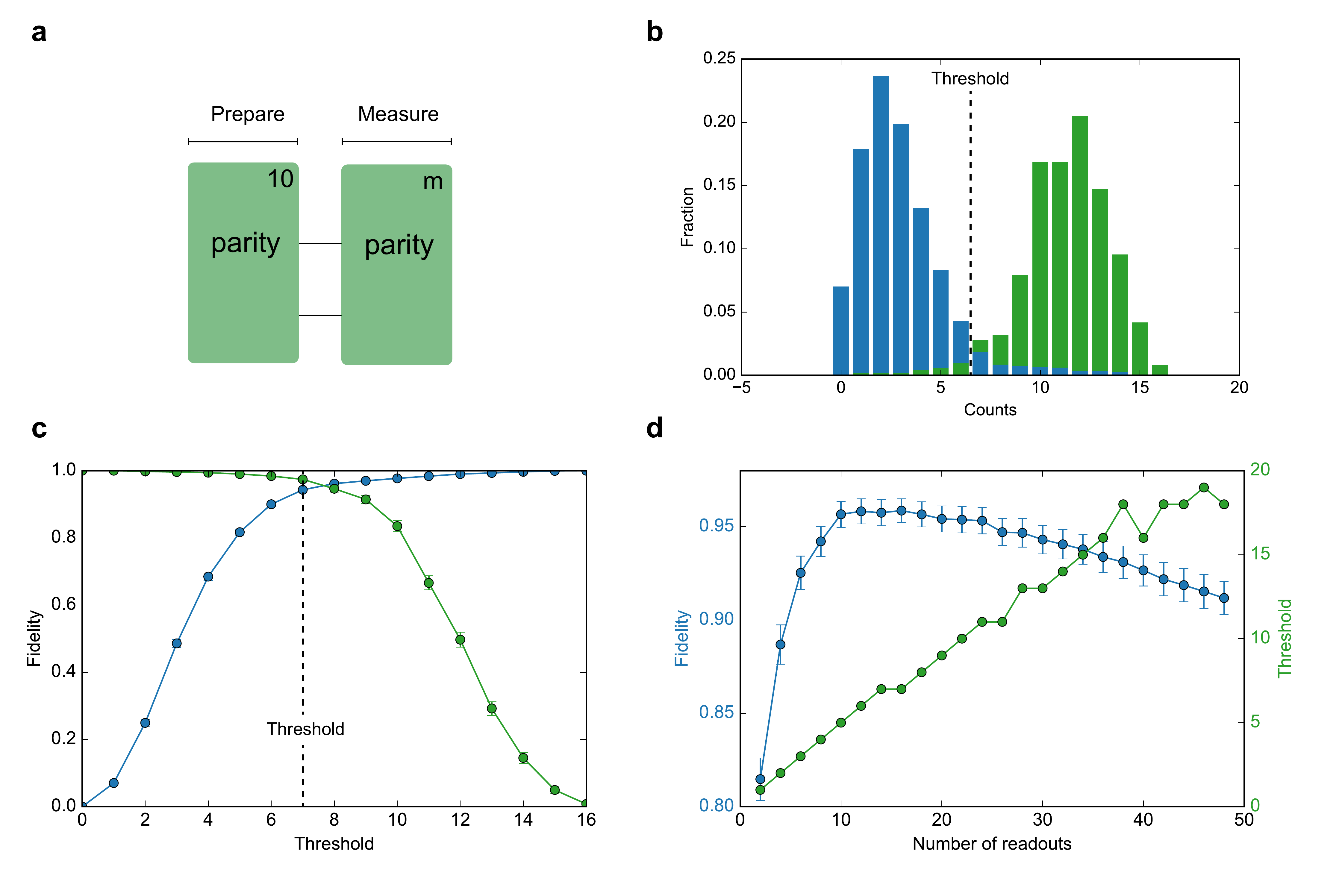}
\caption{\textbf{Parity readout calibration of pair C} \textbf{a.} The sequence used to calibrate the pair C parity readout. We initialise the pair in the antiparallel ($>9/10$) or parallel ($<1/10$) subspace. Then we do $m$ parity readouts that are calibrated to optimally distinguish the subspaces. \textbf{b.} Conditional histograms for the antiparallel (green) and parallel (blue) subspace for $m = 16$ readouts. The optimal threshold of $T = 7$ is indicated. \textbf{c.} Combined initialisation and readout fidelity of the antiparallel (green) and parallel (blue) subspace for $m = 16$ readouts. \textbf{d.} The average fidelity (equation (\ref{fidelity_calculation})) is plotted as a function of the number of readouts $m$ and the corresponding optimal threshold for the given number of readouts.}
\label{pair5_parity_calib}
\end{figure}

\newpage

\section{Value of the dipolar coupling $X$}

The dipolar coupling $X$ between two carbon spins is given by \cite{Zhao2012}
\begin{equation}
    X = \frac{\mu_0 \gamma_c^2 \hbar}{8 \pi r^3}(1-3\cos^2 \theta)
\label{equation_X}
\end{equation}
as defined in Methods, with $\gamma_c = 67.2828 \cdot 10^6 \text{ rad } \text{ s }^{-1} \text{ T }^{-1}$ and $\mu_0 = 4 \pi \cdot 10^{-7}$ H/m. A nearest neighbour pair along the field, such as pair A and B, has $\mathbf{r} = \frac{a_0}{4}[1,1,1]$ where $a_0 = 3.5668$ \r{A} is the lattice constant of diamond at 3.7 K \cite{Stoupin2010}. Consequently $\theta = 0$ and we obtain $X = 2 \pi \cdot 2062.37$ Hz. This is significantly different from the observed value of $X = 2 \pi \cdot 2080.9900(3)$ Hz. For pair C the theoretically predicted value is $X = 2 \pi \cdot 186.92$ Hz whereas the observed value is $X = 2 \pi \cdot 188.33(2)$ Hz. Notably pair A and B have the same or very similar $X$ values, suggesting a mechanism that is not dependent on the local environment. Furthermore, the observed change in $X$ for pair A, B and C is consistent with a reduction of the lattice constant by $\sim 0.01$ \r{A}. In this section, we analyze several mechanisms that could affect the value of $X$.

\subsection{Strain}
The $E_x$ and $E_y$ optical transitions for the NV considered in this work are split by $\sim 4$ GHz, implying a strain of $\delta = 2$ GHz \cite{Abobeih2018}. This is transversal strain with respect to the NV-axis. Axial strain is aligned with the pair A, B axis. We take a typical strain-induced splitting on the order of $\sim 10$ GHz and perform an order of magnitude estimate. 1 GPa of external stress corresponds to a $10^3$ GHz splitting \cite{Davies1976,Batalov2009}, so $\sim 10$ GHz corresponds to $0.01$ GPa. Taking a Young's modulus of $\sim 10^3$ GPa we obtain a deformation of $\sim 10^{-5}$. If the axial strain is comparable to the transversal strain, strain cannot explain the observed increase in $X$, since the required reduction in the lattice constant is on the order of $\sim 10^{-3}$. However, a precise value for the axial strain is not known and we can therefore not rule out this mechanism.

\subsection{Effect of $^{13}$C isotope}
Pair A and B each consist of two $^{13}$C spins surrounded by mostly $^{12}$C spins, because of the natural $^{13}$C abundance of $1.1\%$.  The value for $a_0$ used to get to $X = 2 \pi \cdot 2062$ Hz does not take into account variations in the $^{13}$C abundance. It has been shown that the diamond lattice constant decreases upon increasing the $^{13}$C abundance \cite{Holloway1991}. This suggests that local variations in the diamond lattice constant due to $^{13}$C might play a role in the observed value of $X$. Additional research is required to be able to quantitatively compare this microscopic effect to the measured values.

\subsection{Frequency shifts due to the $^{13}$C bath}
From equation (\ref{motional_narrowing_equation}) we find that the observed frequency in $m_s = 0$ is actually $X+\frac{b^2}{2X}$. The noise contribution $\delta f = \frac{1}{2 \pi}\frac{b^2}{2X}$ is relatively small and therefore we quote the observed value as $X$ in the main text. However, we can estimate the noise term using the values of $b_A = 2 \pi \cdot 13.9(2)$ Hz and $b_B = 2 \pi \cdot 12.5(4)$, obtained from the $T_2^*$ measurement with the NV electron spin in $m_s = -1$. Thus $\delta f_A = 0.046(1)$ Hz and $\delta f_B = 0.038(2)$ Hz. Notably, the difference is $0.009(3)$ Hz. It is therefore possible that the long $T_2^*$ measurements presented in the main text are affected by this frequency difference. However the frequency changes due to the noise $\frac{b^2}{2X}$ are too small to explain the difference between expected and measured value of $X$.

\subsection{Electron-mediated coupling}
\begin{figure}
\includegraphics[scale=0.6]{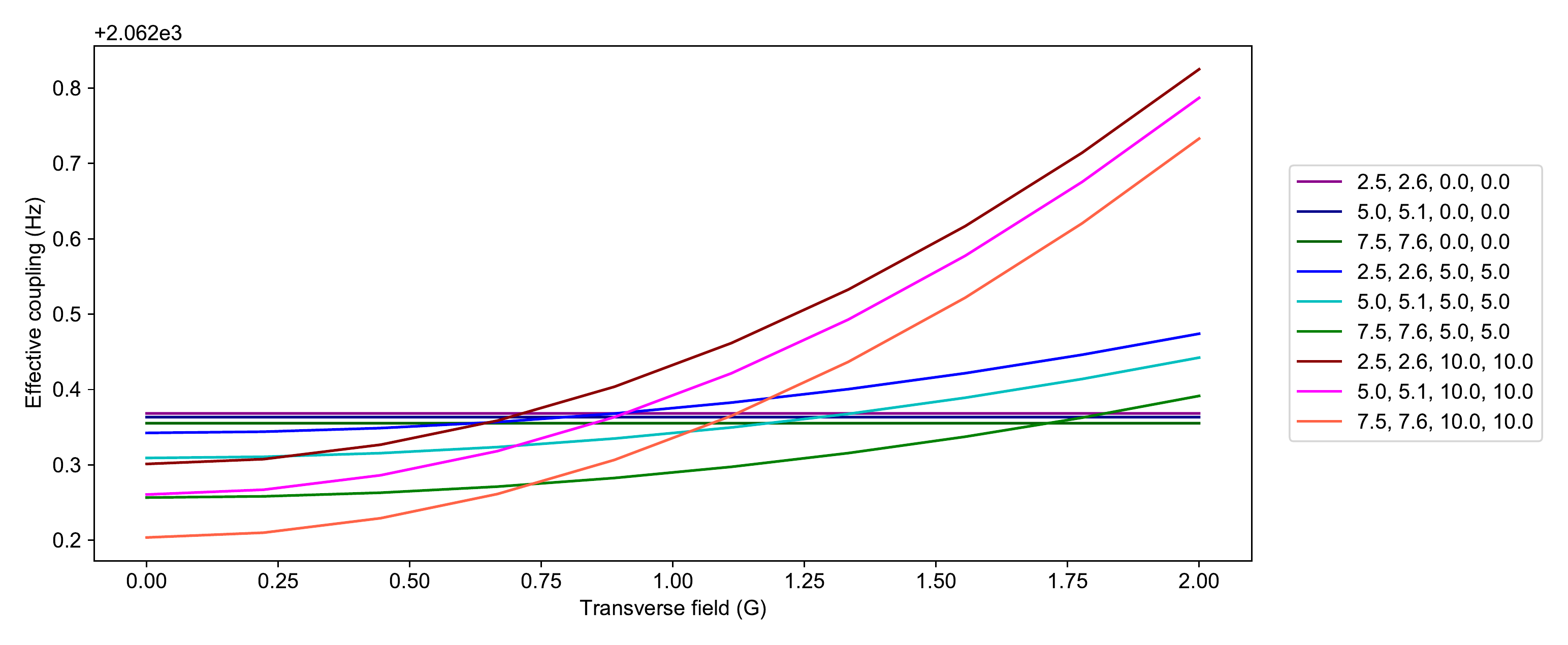}
\centering
\caption{\textbf{Electron-mediated effects on the measured coupling strength $X$.} Numerical simulations showing the obtained effective coupling strength as a function of the transverse magnetic field strength. We show some examples with different combinations of hyperfine couplings between the NV center and the two $^{13}$C spins of the pair. Note that, based on a detailed previous characterization of the nearby environment of this NV center \cite{Abobeih2019} that did not show evidence of these pairs, the actual hyperfine parameters for pairs A and B are expected to be significantly smaller than the largest examples used here. The figure legend shows respectively $A^{(1)}_\parallel$, $A^{(2)}_\parallel$, and $A^{(1)}_\perp$, $A^{(2)}_\perp$ in kHz. For our magnetic field alignment, the transverse field is expected to be well below 1 Gauss \cite{Abobeih2019}, and therefore we conclude that the electron mediated effects in our case are very small ($< 0.5$ Hz). }
\label{Electron mediated effects}
\end{figure}

The coupling strength measured can also be modified due to the presence of the electron spin especially in the presence of a misaligned magnetic field \cite{Abobeih2019}. To study the electron-mediated coupling effects on the measured coupling strength, we consider the Hamiltonian describing a system of an NV center and a pair of coupled $^{13}$C spins with dipolar coupling $X = 2 \pi \cdot 2062.37$ Hz. We numerically solve the full system Hamiltonian (i.e. not considering the pseudo-spin model) to obtain the eigenenergies of the system, from which we calculate the effective coupling strength. Supplementary Figure \ref{Electron mediated effects} shows the obtained effective coupling strength as a function of the transversal magnetic field strength (which reflects how well the field is aligned with the NV axis). We consider some examples with different combinations of hyperfine couplings to the NV center ($A_\parallel$ and $A_\perp$) of the two $^{13}$C spins making up the pair. The maximum values are significantly higher than expected for pairs A and B. For our magnetic field alignment, the perpendicular field is expected to be well below 1 Gauss \cite{Abobeih2019}, and therefore we conclude that the electron mediated effects in our case are very small ($< 0.5$ Hz). Electron-mediated effects are thus very unlikely to explain the difference in the expected and observed value of $X$.

\bibliographystyle{naturemag}
\bibliography{bib}